\documentclass{article}
\usepackage{comment}
\usepackage{amsfonts}
\usepackage{amsthm}
\usepackage{bbding}
\usepackage{pifont}
\usepackage{nicematrix}
\usepackage[english]{babel}

\usepackage[letterpaper,top=2cm,bottom=2cm,left=3cm,right=3cm,marginparwidth=1.75cm]{geometry}

\usepackage{amsmath}
\usepackage{graphicx}
\usepackage[colorlinks=true, allcolors=blue]{hyperref}
\usepackage{bm}

\usepackage{biblatex}
\title{A Class of Markovian Self-Reinforcing Processes with Power-Law Distributions}

\author{
  Pavlo Bulanchuk
  \thanks{Corresponding authors:\\
  Pavlo Bulanchuk: bulanchukp@hhmi.org\\
  Sue Ann Koay: koays@hhmi.org\\
  Sandro Romani: romanis@hhmi.org},
  Sue Ann Koay\footnotemark[1],
  Sandro Romani\footnotemark[1] \\
  {\small Janelia Research Campus, Howard Hughes Medical Institute, Ashburn, VA, USA}
}

\begin{document}
\maketitle

\begin{abstract}
Solar flares, email exchanges, and many natural or social systems exhibit bursty dynamics, with periods of intense activity separated by long inactivity. These patterns often follow power-law distributions in inter-event intervals or event rates. Existing models typically capture only one of these features and rely on non-local memory, which complicates analysis and mechanistic interpretation. We introduce a novel self-reinforcing point process whose event rates are governed by local, Markovian nonlinear dynamics and post-event resets. The model generates power-law tails for both inter-event intervals and event rates over a broad range of exponents observed empirically across natural and human phenomena. Compared to non-local models such as Hawkes processes, our approach is mechanistically simpler, highly analytically tractable, and also easier to simulate. 
We provide methods for model fitting and validation, establishing this framework as a versatile foundation for the study of bursty phenomena.

\end{abstract}

\section{Introduction}

Bursty dynamics - periods of high activity separated by periods of prolonged quiescence - are pervasive in natural and behavioral systems. Such intermittent behavior is empirically observed across a broad range of phenomena, including financial market trades \cite{gabaix_theory_2003}, earthquakes \cite{corral_long-term_2004, touati_origin_2009}, forest fires \cite{malamud_forest_1998}, neuronal avalanches \cite{beggs_neuronal_2003}, human communication patterns \cite{karsai_universal_2012}, and free-choice behavior in both humans \cite{henderson_modelling_2001} and animals \cite{sorribes_origin_2011, proekt_scale_2012}. A common feature of these bursty processes is the presence of heavy-tailed distributions for either inter-event intervals (IEIs) or instantaneous event rates (also often referred to as intensities), which are often well-described by power laws (see Table S1 for details).
The appeal of describing long-tailed distributions with simple functions, such as power laws, arises from an intuitive assumption that there should be correspondingly simple and general generative mechanisms; yet, despite extensive empirical documentation, such mechanisms are often lacking. 

Several classes of models have been proposed to account for heavy-tailed IEI distributions (Table S2). Priority queuing mechanisms, where tasks with random priorities arrive at a constant rate, successfully describe burstiness in human communication \cite{barabasi_origin_2005, Vazques2006}. Cox-like processes, where event emissions are driven by autonomous fluctuations (either stochastic \cite{kleinberg_bursty_nodate} or seasonal \cite{Malmgren2008}) of the underlying event rates, have also been proposed as general frameworks to capture long-tailed IEI distributions. Hawkes processes, characterized by self-excitation through non-local memory kernels, represent another major class of models.  However, standard Hawkes processes inevitably produce exponential cutoffs in IEI distributions due to the presence of a constant background rate. Genuine power-law tails can nonetheless be achieved in extensions that impose specific constraints on the process's kernel and memory and have a zero background rate \cite{Jo2015}. Additionally, burstiness in phenomena such as solar flare occurrences has often been modeled using self-organized criticality (SOC) frameworks, where power-law statistics arise from fluctuations in the energy input rate statistics. \cite{Bak1998, Norman2001, Wheatland2000}.

Complementarily, several models have been proposed for explaining power laws in the underlying instantaneous event rates. \textit{Nonlinear} Hawkes processes can generate a power law in rates when either close to criticality \cite{kanazawa_asymptotic_2023}, or under certain constraints \cite{kanazawa_ubiquitous_2021}. Langevin equations incorporating multiplicative noise also inherently produce power-law distributions \cite{Biro2005}. In mobile text messaging, a two-state self-reinforcing model explains the power-law distribution observed for consecutive message counts (a quantity related, although not identical, to event rates) \cite{karsai_universal_2012}. In financial markets, the observed power-law distributions for daily trade frequencies have been modeled with Pareto-distributed sizes of trading companies combined with optimization-driven trade size selection \cite{plerou2000}.

The models mentioned above have several conceptual and practical limitations. Cox-type models lack any feedback from the event occurrences onto the underlying event rates, which is limiting. Domain-specific approaches --- such as priority queue models for human communication, energy input fluctuations in SOC models for solar flares, or models relating trade frequencies to Pareto-distributed trading company sizes --- are highly context-specific and require construction of a distinct model for each phenomenon. When applying the Hawkes framework to capture the power-law statistics of bursty phenomena, the variants most commonly used in the literature employ non-local, long-memory kernels \cite{Sornette_2006, Sornette_2007, Bacry_2015}. 
While effective, this non-locality complicates mechanistic interpretation and still features exponential cutoffs. Non-linear Hawkes processes with exponential kernels can overcome these problems by the appropriate choice of non-linearity, and they can also be reformulated in a local form, but the structure of the model is rather constraining (more on that in the Discussion section). The constraints and complexities of these existing approaches highlight a need for a more parsimonious class of Markovian models, where power-law distributions emerge as a natural consequence of simple, intrinsic dynamics across a wide parameter range.


We address the above gap by introducing a local (Markovian) self-excited point process, the intensity of which decays deterministically in between events and is reset by a straightforward rule after each event. Across a wide (analytically derived) regime of parameter space, this parsimonious framework can generate power-law distributions in both IEIs and event rates without requiring non-local memory kernels or stochastic rate modulations. Our model provides an analytically tractable, computationally efficient, and broadly applicable approach to modeling bursty dynamics.

The remainder of the paper is organized as follows: In Section 2, we provide an intuition for why a decay term proportional to $\lambda^2$ naturally leads to power-law-distributed waiting times. We then introduce three specific model variants: one with a constant post-event reset ("constant reset model"), another with a linear intensity reset ("linear jump scaling model"), and the third scale invariant model. We analyze their steady-state intensity distributions, mean event rates, and correlation structures. We finish the chapter by developing intuition for a general form of a reset function. Section 3 describes the procedures for simulating these models and fitting their parameters to empirical data using maximum likelihood estimation, as well as validation of the model. Finally, in Section 4, we summarize our findings and discuss potential applications and future extensions. Additional technical derivations and supporting analyses are presented in the appendices.

\section{Theoretical Framework}
\subsection{General Intuition for Power-Law Behavior}
To understand how a power-law distribution for inter-event intervals (IEIs) can emerge, we start by establishing a relationship between the intensity (instantaneous rate) of the process $\lambda(t)$, and the IEI survival function $S(\tau)$. 

Assume that an event has occurred at $t = 0$, and that the intensity $\lambda(t)$ evolves deterministically until the next event. The probability that no subsequent event occurs in an infinitesimal interval $[t, t+\mathrm{d}t]$ is $1 - \lambda(t)\mathrm{d}t$. Then, the probability that waiting time $T$ for the next event will be greater than $\tau$  is the product of the probabilities of having zero events at each time step:

\begin{equation}
\begin{split}
     P(T > \tau) &= \left(1-\lambda(0)dt\right)...(1-\lambda(\tau)dt) \\
    &= \exp\left(-\int_0^\tau \lambda(t)\,\mathrm{d}t\right).\,
\end{split}
    \label{eq:survival_function}
\end{equation}
This defines the survival function $S(\tau) \equiv P(T > \tau)$ of inter-event intervals. By differentiating the logarithm of this survival function, we obtain the intensity as a function of the survival function:
\begin{equation}
    \lambda(\tau) = -\frac{\mathrm{d}}{\mathrm{d}\tau}\ln S(\tau).
    \label{eq:lambda_from_survival}
\end{equation}
We can now reverse engineer the dynamics required for $\lambda(t)$. If we postulate a power-law form for the IEI probability density function (PDF), $p(\tau) = -\mathrm{d}S(\tau)/\mathrm{d}\tau \sim \tau^{-\alpha}$, then the corresponding survival function must scale as $S(\tau)\sim \tau^{-\alpha+1}$. Substituting this into Eq. (\ref{eq:lambda_from_survival}) yields the necessary form for the intensity:
\begin{equation}
    \lambda(\tau) \sim \frac{\alpha-1}{\tau}.
    \label{eq:lambda_scaling}
\end{equation}
By taking the time derivative of Eq. \eqref{eq:lambda_scaling}, we obtain $\dot{\lambda} \propto -1/\tau^2$, which suggests that a differential equation with $\dot{\lambda} \propto -\lambda^2$ should produce the desired scaling. It is straightforward to demonstrate that if $\dot{\lambda}$ decays more slowly (e.g., $\dot{\lambda} \propto -\lambda^p$ with $p > 2$), the IEIs decay faster than a power law. If, instead, $p < 2$, the process eventually ceases activity due to the rapid decay of $\lambda$.  This thus motivates the central dynamical rule of our model: $\dot{\lambda} = -a\lambda^2$.

\subsection{General class of considered models}

Building on the intuition from the previous section, we now define the general class of Markovian self-reinforcing processes considered in this paper. The dynamics of the process is governed by two rules:
\begin{enumerate}
    \item \textbf{Inter-event decay.} Between two consecutive events, the intensity $\lambda(t)$ evolves deterministically, according to the quadratic decay rule:
    \begin{equation}
    \dot{\lambda} = -a\lambda^2,
    \end{equation}
where $a$ is a dimensionless constant we term the {\it decay parameter}.
    \item \textbf{Post-event reset.} Upon the occurrence of an event, the intensity is instantaneously reset from its value just prior to the event, $\lambda^-$, to a new value immediately after, $\lambda^+$. This jump is determined by a reset function
\begin{equation}
    \lambda^+ = f(\lambda^-).
\end{equation}
\end{enumerate}

Although our framework can accommodate any reset function, this paper will focus on the linear case, which, though simple, is sufficiently rich to generate a wide array of dynamics.

\subsection{Constant reset model}

We begin by analyzing the simplest instantiation of the general framework: the \textbf{constant reset model}. In this case, the reset function is constant, $f(\lambda) = \lambda_0$. Because the intensity is reset to the same value $\lambda_0$ after every event, this model is a particular case of a renewal process \cite{cox_renewal_1967},  which is defined by independently drawn inter-event times. 

The intensity dynamics can be expressed in the form of a stochastic differential equation:
\begin{equation}
    \mathrm{d}\lambda = -a\lambda^2\mathrm{d}t + (\lambda_0 - \lambda)\eta,
    \label{eq:baseline}
\end{equation}
where $\eta \sim \mathrm{Ber}(\lambda\mathrm{d}t)$ is the event indicator, equal to 1 when an event occurs and 0 otherwise (see Fig. 1a for an example simulation).

\begin{figure}
\centering
\includegraphics[width=1.\linewidth]{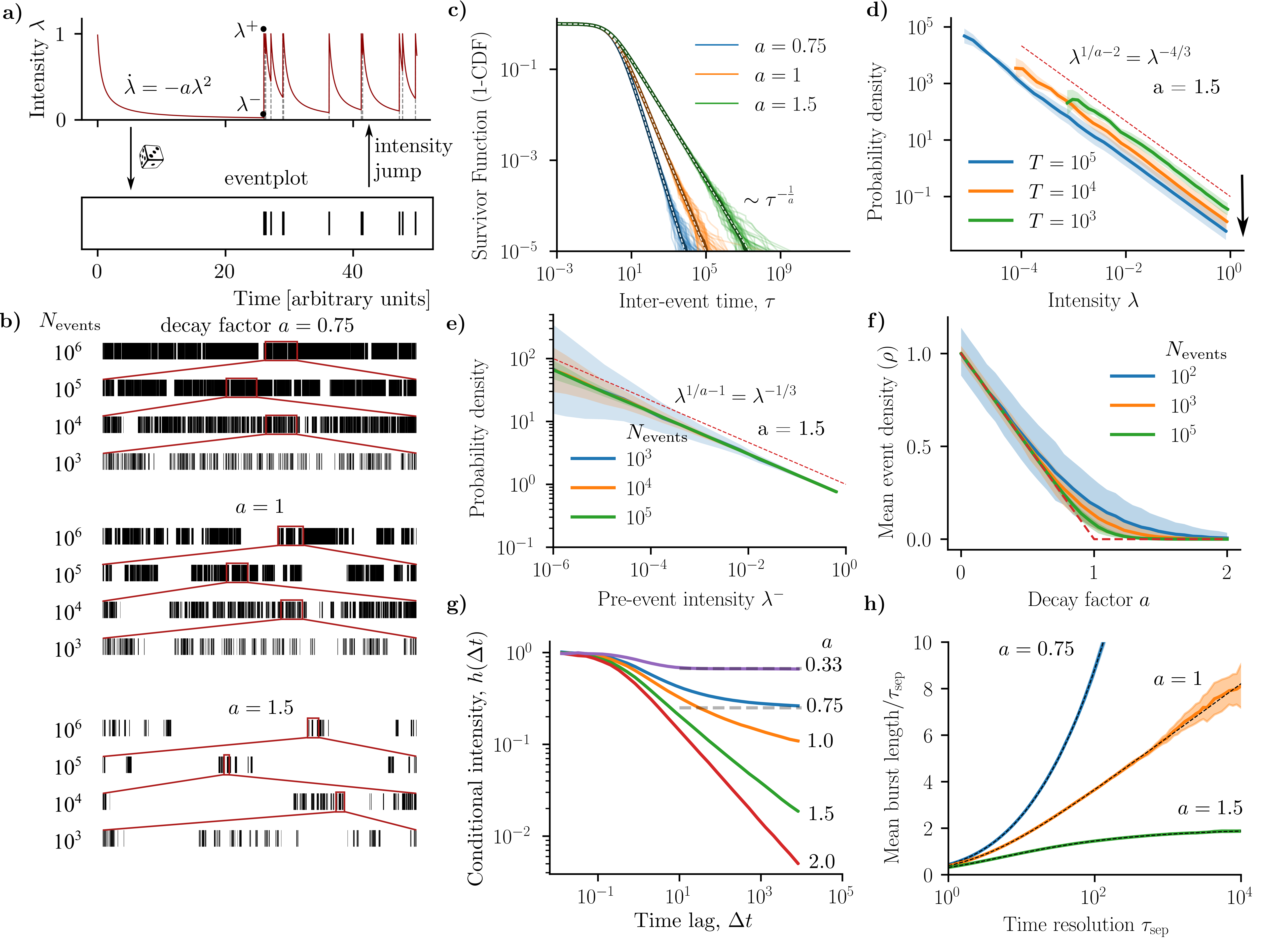}
\caption{Characteristics of the stochastic process with constant reset, defined by Eq.~\eqref{eq:baseline}. 
a) Illustration of a simulation run with parameters $a=1.5$ and $\lambda_0 = 1$. 
b) Event plots for different values of the decay parameter $a$ and total number of simulated events. 
c) Simulated survival function for the inter-event interval (IEI) distribution. Dashed lines correspond to the theoretical stationary distributions.
 d) Empirical probability density distribution of the intensity $\lambda$ for decay factor $a = 1.5$ and different simulation durations. The probability density follows the theoretical power-law of $\lambda^{-4/3}$, but since the distribution is non-normalizable, the curve shifts downwards as the simulation length increases. 
 e) Empirical probability density distribution of the pre-event intensity $\lambda^-$ for decay factor $a = 1.5$ and different numbers of simulated events. 
 f) Mean event density ($\rho$) as a function of the decay parameter $a$.  The dashed line shows the exact analytical solution; other lines indicate the simulated median $\rho$ for different total numbers of events. Shaded areas represent the 10th–90th percentile range.
 g) The conditional intensity of observing an event at time $t+\Delta t$ given an event at $t$, $p(\text{event at } t+\Delta t| \text{event at }t)$, as a function of the time lag $\Delta t$.
 h) The mean duration of a continuous burst of events (as seen in panel Fig 1b) vs. the minimal gap resolution $\tau_{\text{res}}$ used to define it. 
 }
\label{fig:1_constant_start}
\end{figure}

Even in such a simple model, the resulting event plots exhibit fractal-like structure (Fig. \ref{fig:1_constant_start}b), a feature that, as we will show, is a consequence of the power-law distribution for IEIs.

\subsubsection{Key Characteristics of the process}
\paragraph{Inter-event interval distribution.}
Let us assume that the last event happened at $t = 0$, that is, $\lambda(0)=\lambda_0$. Solving the decay equation, $\dot{\lambda} = -a \lambda^2$ with this initial condition gives the intensity at a later time (before the next event): 
\begin{equation}
    \lambda(t) = \frac{\lambda_0}{1 + \lambda_0at}.
    \label{eq:lambda_sol}
\end{equation}
By substituting this into the general formula for the survival function (\ref{eq:survival_function}), we find:
\begin{equation}
    S(\tau)\equiv P(t > \tau) = \left(1 + \lambda_0a\tau\right)^{-1/a},
    \label{eq:CDF1}
\end{equation}
and differentiating this expression ($p(\tau) = -\mathrm{d}S/\mathrm{d}\tau$) yields the corresponding probability density:
\begin{equation}
    p(\tau) = \lambda_0\left(1 + \lambda_0a\tau\right)^{-1/a - 1}.
    \label{eq:inter_event1}
\end{equation}
For large $\tau$, this PDF exhibits a power-law tail, $p(\tau)\propto \tau^{-\alpha_{\text{IEI}}}$, with an exponent $\alpha_{\text{IEI}}=1 + 1/a$. Fig. 1c presents the survival function from Eq.~(\ref{eq:CDF1}) alongside simulated data for individual simulations, which are closely matched in the bulk of the distributions up to expected stochastic deviations in the tail. 

\paragraph{Stationary intensity distribution}
An important feature of the processes considered in this paper is the possibility to have non-normalizable stationary distributions, such as the stationary intensity distribution $p(\lambda)$. To find its functional form, we use the stationary master equation, which is valid for $\lambda < \lambda_0$ (see Appendix \ref{ap:master} for a derivation):
\begin{equation}
    \frac{\partial}{\partial\lambda}\left(a\lambda^2p(\lambda)\right) - \lambda p(\lambda) = 0,
    \label{eq:fokker0}
\end{equation}
This equation describes a steady-state balance.The first term represents the divergence of the probability flux driven by the deterministic decay of $\lambda$. The second term, $-\lambda p(\lambda)$, acts as sink, accounting for the loss of probability density from state $\lambda$ when an event triggers a reset. Because all post-event resets in this model end up at the specific value $\lambda = \lambda_0$, there is no corresponding source term for any state $\lambda < \lambda_0$. Solving this differential equation yields a power-law form:
\begin{equation}
    p(\lambda) \propto \lambda^{1/a - 2}.
    \label{eq:non-norm}
\end{equation}
For this distribution to be normalizable, its integral must converge, which requires the exponent to be greater than -1. The condition $1/a - 2 > -1$ holds only for $a <1$. For $a \geq 1$ the distribution becomes non-normalizable (because of its behavior at small $\lambda$), implying that no proper stationary distribution exist.

The absence of a stationary distribution for $a\ge 1$ may initially seem counterintuitive, given that the process remains  recurrent (the probability of an infinite waiting time is zero, as seen from Eq. (\ref{eq:CDF1})). This apparent paradox is resolved by noting that for $a \ge 1$ the process is \textbf{null-recurrent}: although its return to any given state is certain, the mean time to return is infinite since the mean IEI diverges for $a\ge 1$.  

 A more complete understanding of the $a\ge 1$ regime emerges from \textit{finite-time} simulations, which exhibit the expected power-law behavior and exponent (\ref{eq:non-norm}), but with a cutoff at small $\lambda$. As the simulation duration increases, this cutoff shifts toward smaller $\lambda$ values.  Consequently, since the total integral must be preserved, the overall scaling of the distribution goes down (see Fig. 1d).  This effect is an example of an intermediate asymptotics.
 
 Interestingly, even when the stationary $p(\lambda)$ is ill-defined, assuming the form $p(\lambda)=c\lambda^{1/a-2}$ allows us to derive the pre-event intensity distribution $\lambda^-$ (intensity right before an event):
\begin{equation}
    p^-(\lambda^-)\propto \lambda^-  p(\lambda^-)\propto\left(\lambda^-\right)^{1/a - 1},
    \label{eq:norm}
\end{equation}
which is normalizable for any value $a>0$. As shown in Figure 1e, simulations confirm this predicted form.

\paragraph{Mean Density.} The mean density of events for a simulation of length $T$ that contains $N$ events is defined as $\rho \equiv N/T$, which is the reciprocal of the mean inter-event time $\rho = 1/\langle\tau\rangle$. The value of $\langle \tau\rangle$ is found by integrating the survival function, Eq.~(\ref{eq:CDF1}). This integral converges only for $a < 1$, yielding a mean density of:
\begin{equation}
    \rho = (1-a)\lambda_0.
    \label{eq:mean}
\end{equation}
For $a \geq 1$, the mean IEI diverges, and consequently, the mean event density is zero.

\paragraph{Correlational Structure.} To quantify temporal correlations between events, we use the conditional intensity $h(\Delta t)\equiv p(\textrm{event at } t +\Delta t| \textrm{event at }t)$ (where $p$ here is a probability per unit of time).
Unlike the autocorrelation function $R(\Delta t)=p(\textrm{event at } t+\Delta t, \textrm{event at }t)$, which vanishes for zero-mean processes such as those with $a>1$, the conditional probability remains informative for all values of $a$. The two quantities differ only by a factor $p(\textrm{event at } t)=\rho$. Figure 1g shows that for this model correlations decay as a power-law for $a>1$, reproducing the power law tails of a similar renewal process model in \cite{vajna_modelling_2013}.

\paragraph{Fractal-like structure of event plots.} A remarkable feature of event plots from this model is their fractal-like bursty structure for some values of $a$ (Fig. 1b). This structure can be quantified by how the length of event "bursts" scale with the time resolution used to identify bursts, $\tau_{\mathrm{sep}}$. Formally, we define a \textbf{burst} as a sequence of events where the inter-event gaps are all smaller than a chosen $\tau_{\mathrm{sep}}$. The mean burst duration $l(\tau_{\mathrm{sep}})$ can then be derived from the IEI survival function (see Appendix \ref{ap:fractal}):
\begin{equation}
    l(\tau_{\mathrm{sep}}) = S(\tau_{\mathrm{sep}})^{-1} \int_0^{\tau_{\mathrm{sep}}} S(t) dt - \tau_{\mathrm{sep}},
\end{equation}
For our model with $\lambda_0 = 1$, this simplifies to:

\begin{equation}
l(\tau_{\mathrm{sep}}) = 
\begin{cases}
\displaystyle\frac{1 + \tau_{\mathrm{sep}} - (1 + a\tau_{\mathrm{sep}})^{1/a}}{a-1} 
& \text{if } a \neq 1, \\
(1+\tau_{\mathrm{sep}})\ln(1+\tau_{\mathrm{sep}}) - \tau_{\mathrm{sep}} 
& \text{if } a = 1.
\end{cases}
\end{equation}

The asymptotic behavior of $l(\tau_{\mathrm{sep}})$ for large $\tau_{\mathrm{sep}}$ reveals three distinct scaling regimes governed by the decay parameter $a$:
\begin{itemize}
    \item $a > 1$ : The mean duration scales linearly, $l(\tau_{\mathrm{sep}}) \propto \tau_{\mathrm{sep}}$. Here, the relative burst duration $l(\tau_{\mathrm{sep}})/\tau_{\mathrm{sep}} $ approaches a constant, hence the process appears statistically self-similar at large scales (Fig. 1b, bottom).
    \item $a = 1$ : The scaling is nearly linear, $l(\tau_{\mathrm{sep}}) \sim \tau_{\mathrm{sep}} \ln \tau_{\mathrm{sep}}$. The relative duration grows logarithmically, so bursts become more prominent as one "zooms out" (Fig. 1b, middle).
    \item $a < 1$ : The duration grows faster than linearly, $l(\tau_{\mathrm{sep}}) \propto \tau_{\mathrm{sep}}^{1/a}$. The relative duration increases as a power law, $\tau_{\mathrm{sep}}^{1/a-1}$, meaning bursts rapidly aggregate and dominate the timeline (Fig. 1b, top).
\end{itemize}

Figure~1h illustrates $l(\tau_{\mathrm{sep}}) \sim \tau_{\mathrm{sep}}$ for these three scaling regimes, showing how $\tau$ controls the aggregation of events into blocks.

\subsection{Linear Jump Scaling model}
In the constant reset model, the post-event intensity is always fixed at $\lambda_0$. This means the process "forgets" its state after each event, making it a renewal process with uncorrelated inter-event intervals (IEIs). To introduce memory effects via a \textit{local} mechanism, we generalize the model by allowing the post-event intensity to scale linearly with the pre-event intensity:
\begin{equation}
    \lambda_i^+ = k\lambda_i^- + c.
    \label{eq:iterative}
\end{equation}
Here, the dimensionless \textit{gain parameter} $k$  controls the scaling of the feedback, and $c$ is a constant offset with units of $\left[T^{-1}\right]$. This more general formulation includes the constant reset model as a special case, recovered by setting $k = 0$ and $c = \lambda_0$. Some examples of the resulting dynamics are shown in Fig. 2a.

The entire process can be described by a single stochastic differential equation (SDE): 
\begin{equation}
    \mathrm{d}\lambda = -a\lambda^2\mathrm{d}t + \left[(k-1)\lambda + c\right]\eta,
    \label{eq:general}
\end{equation}
where, as before,  $\eta \sim \text{Ber}(\lambda dt)$ is the event indicator. 

Equation~(\ref{eq:general}) can be reduced to dimensionless form by an appropriate scaling of the time axes, leading to $c = 1$ (see Appendix \ref{ap:scaling}). Thus, we set $c = 1$ in all subsequent simulations for simplicity.

\paragraph{Event-Domain Representation}
While the SDE provides a continuous-time description of the intensity dynamics in the \textit{time domain}, it is often more insightful and analytically convenient to analyze the process in the discrete \textit{event domain}. 
This framework maps the intensity from one event to the next, effectively integrating out the inter-event dynamics. 

The evolution from event $i$ to event $i+1$ occurs in two steps. First, at event $i$, the intensity deterministically jumps from its pre-event value $\lambda^-_i$ to the post-event value $\lambda^+_i$ according to the model's reset rule \eqref{eq:iterative}. Second, during the subsequent random waiting time $\tau_i$, the intensity stochastically decays  from $\lambda^+_i$ to the next pre-event value, $\lambda_{i+1}^-$. We can derive the cumulative distribution for the $\lambda_{i+1}^-$, $F(\lambda_{i+1}^-|\lambda_i^+)\equiv P(\text{pre-event intensity }< \lambda_{i+1}^-|\lambda_i^+)$ by noting that the probability of waiting time exceeding $\tau$ (the survival function $S(\tau|\lambda_i^+)$) is  is equivalent to the probability of the intensity decaying below the corresponding value $\lambda^-_{i+1}(\tau)$. Using the decay rule $\dot{\lambda} = - a\lambda^2$, we can re-express the integral over time as an integral over intensity:
\begin{equation} 
\begin{split}
    F(\lambda_{i+1}^-|\lambda_i^+) &= S(\tau)= \exp\left(-\int_0^\tau \lambda(t) dt\right)\\
    & = \exp\left(-\int_{\lambda_i^+}^{\lambda_{i+1}^-} \frac{\lambda d\lambda}{\dot{\lambda}}\right)\\
    & = \exp\left(\int_{\lambda_i^+}^{\lambda_{i+1}^-} \frac{ d\lambda}{a\lambda}\right)\\
    &= \left(\frac{\lambda_{i+1}^-}{\lambda_i^+}\right)^{1/a}
\end{split}
\label{eq:iterative_CDF}
\end{equation}
Differentiating this CDF with respect to $\lambda^-_{i+1}$  yields the probability distribution
\begin{equation}
    p(\lambda^-_{i+1}|\lambda^+_i) = \frac{1}{a \lambda^-_{i+1}}\left(\frac{\lambda^-_{i+1}}{\lambda^+_i}\right)^{1/a}.
    \label{eq:iterative_prob_distrib}
\end{equation}
This result can be reformulated in a different generative form via inverse transform sampling.  Setting the CDF \eqref{eq:iterative_CDF} equal to a uniform random variable $U_i\sim \mathrm{Unif}(0,1)$ we can generate the next pre-event intensity as :
\begin{equation}
\lambda_{i+1}^- \;=\; F^{-1}(U_i\mid \lambda_i^+) \;=\; \lambda_i^+\,U_i^{\,a}.
\label{eq:sampling_identity}
\end{equation}
Combining this with the linear reset rule $\lambda_{i+1}^+=k\,\lambda_{i+1}^-+c$, yields the random-coefficient AR(1) recursion
\begin{equation}
\lambda_{i+1}^+ \;=\; (k\,U_i^{\,a})\,\lambda_i^+ + c,
\qquad U_i \stackrel{\text{i.i.d.}}{\sim}\mathrm{Unif}(0,1),
\label{eq:kesten_recursion}
\end{equation}
which is specific instance of a Kesten process \cite{kesten1973}
\begin{equation}
    X_{i+1}=A_i X_i+B_i
    \label{eq:kesten_recursion_basic}
\end{equation}
with coefficients $A_i=kU_i^{a}$ and $B_i=c$. The iterative equation for the pre-event intensity is also a Kesten process:
\begin{equation}
    \lambda_{i+1}^- \;=\; (k\,U_i^{\,a})\,\lambda_i^- + c U_i^{\,a} .
    \label{eq:kesten_recursion__}
\end{equation}

\begin{figure}
\centering
\includegraphics[width=1.\linewidth]{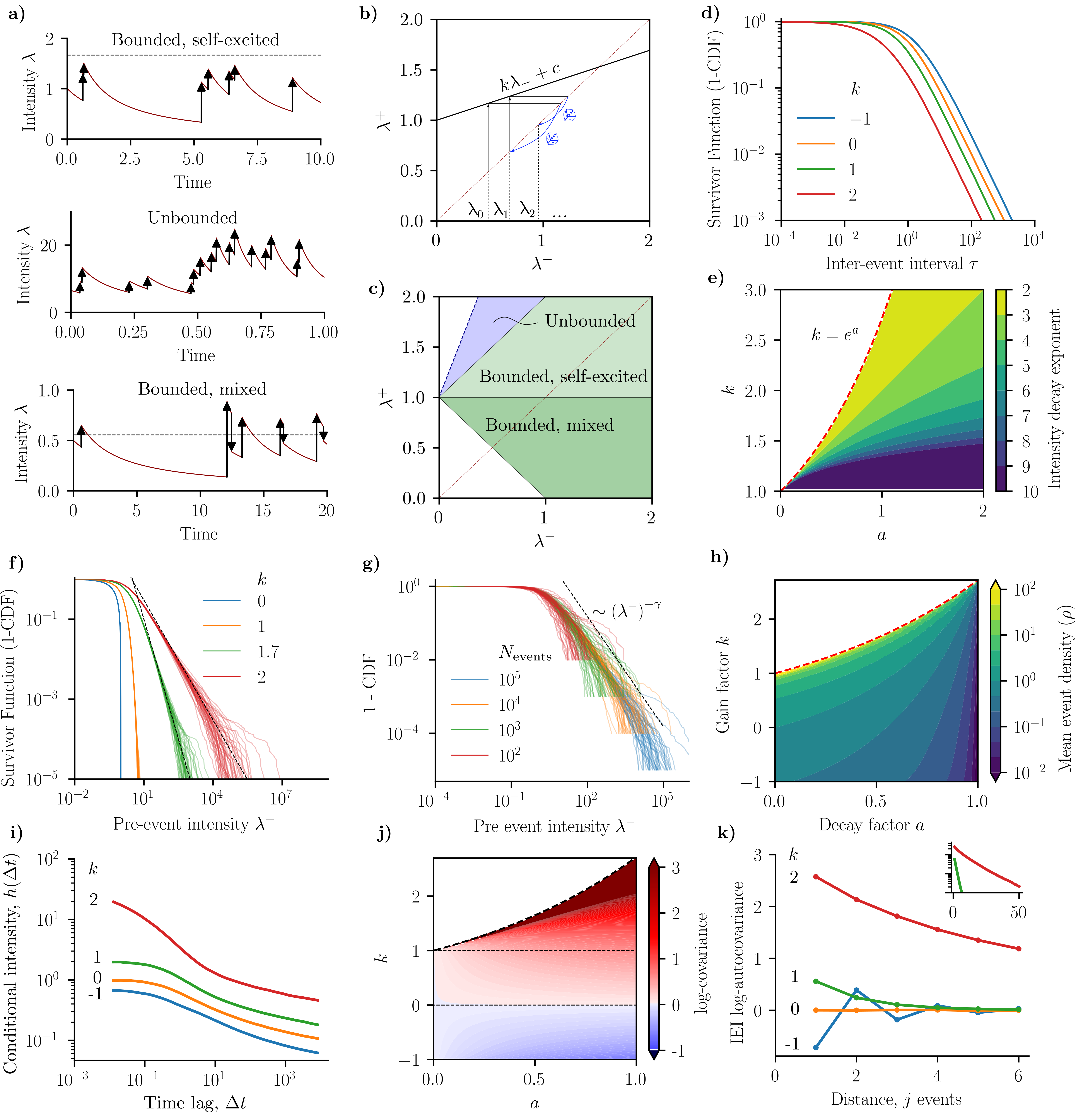}
\caption{Characteristics of the process with linear reset function. 
a) Example simulations of the intensity evolution for $a=0.5, c=1$, and three different values of the gain parameter $k$, corresponding to three dynamical regimes. Top: Bounded, self-excited ($k=2/3$). Middle: Unbounded ($k=1.5$). Bottom: Bounded, mixed-feedback ($k=-0.8$).
b) An illustration of the event-to-event intensity map, $\lambda^+ = k\lambda^- + c$. The cobweb plot shows how the pre-event ($\lambda^-$) and post-event ($\lambda^+$) intensities evolve over successive events.
c) The parameter space of the model, showing the unbounded, bounded self-excited, and bounded mixed-feedback regimes as a function of the gain parameter $k$
d) The survival function of inter-event intervals for $a=1$ and different values of $k$. The identical power-law tail demonstrates that the IEI exponent is independent of $k$
e) The power-law exponent of the stationary intensity distribution in the limit $\lambda \rightarrow \infty$. The red dashed line delineates the stability threshold beyond which the model becomes unstable.
f) Empirical survival function of the pre-event intensity $\lambda^-$ for different values of the gain parameter $k$ and $a = 1$. The dashed lines correspond to the theoretical power-law trend for the stationary distribution.
g) Empirical survival function of the pre-event intensity for different simulation lengths, $a = 1$ and $k = 2$.
h) The mean event density, $\rho$, as a function of $a$ and $k$, calculated from simulations.
i) The conditional intensity $h(\Delta t)\equiv p(\textrm{event at } t+\Delta t| \textrm{event at } t)$, for $a=1$ and different values of $k$.
j) Covariance of the logarithm for neighboring inter-event intervals as a function of the model parameters $a$ and $k$. 
k) Covariance of the logarithm for inter-event intervals with $j$ events between them for $a = 1$.
}
\label{fig:2}
\end{figure}

\paragraph{Parameter Regimes}
The behavior of the model is qualitatively determined by the gain parameter $k$, which divides the parameter space into distinct regimes:

\begin{enumerate}
    \item \textbf{Bounded processes ($-1\le k < 1$)}: When the gain is less than one, the  process is strictly bounded from above. The intensity eventually enters and remains confined by the maximum possible intensity value $\lambda_{\text{max}} = c/(1-k)$. This class of stable processes can be further divided into two subclasses based on the sign of  $k$:
    \begin{itemize}
        \item Self-excited ($k > 0$): The intensity jump after an event is always positive. 
        \item Mixed ($k<0$): The intensity jump can be both positive and negative.
    \end{itemize}
    As we will show, these subclasses also differ in their temporal correlation structure: self-excited processes exhibit positive correlations between neighboring IEIs, while the mixed processes exhibit negative correlations. 
    
    \item \textbf{Unbounded processes ($k \geq 1$)}: When the gain is one or greater, the intensity is no longer strictly bounded and can reach arbitrarily large values. However, the process is not necessarily unstable: it is \textbf{recurrent} for $k < e^a$, and \textbf{transient} for $k > e^a$. A useful intuition about the threshold $k = e^a$ can be gained from the scale invariant process ($c = 0$), which is discussed in Section~\ref{sec:scale_invariant_model}, and the rigorous result can be established from the Kesten form, Eq. \eqref{eq:kesten_recursion} (see Appendix \ref{ap:stability_analysis}).
\end{enumerate}

\paragraph{Inter-event interval distribution.} Since $\lambda_+$ is no longer a constant, the probability distribution for IEIs is now a mixture over the possible values of $\lambda^+$:
\begin{equation}
    p(\tau) = \int_{0}^{\infty} p(\tau|\lambda^+)p(\lambda^+)d\lambda^+.
    \label{eq:inter_event_general_1}
\end{equation}
The conditional probability $p(\tau|\lambda^+)$ retains the form of the constant reset model (\ref{eq:inter_event1}), but with $\lambda^+$ instead of $\lambda_0$:
\begin{equation}
    p(\tau|\lambda^+) = \lambda^+\left(1 + \lambda^+ a\tau\right)^{-1/a - 1}.
    \label{eq:inter_event2}
\end{equation}
Crucially, integrating over $\lambda^+$ preserves the power-law tail, which remains $p(\tau)\sim \tau^{-1-1/a}$ and is therefore independent of the value of $k$. This is demonstrated numerically in Fig. 2d and is derived formally in  Appendix \ref{ap:iei_prob_distrib}.

\paragraph{Stationary intensity distribution.}
The stationary distribution of $\lambda$ can be obtained from the corresponding master equation derived in Appendix~\ref{ap:master}:
\begin{equation}
    \frac{\partial}{\partial\lambda}\left(a\lambda^2p(\lambda)\right) - \lambda p(\lambda) + \frac{1}{k}p\left(\frac{\lambda - c}{k}\right)\frac{\lambda - c}{k} = 0.
    \label{eq:fokker2}
\end{equation}
For unbounded processes ($k > 1$), the stationary distribution exhibits a power-law tail. To find its exponent, we consider the asymptotic regime $\lambda \to \infty$, where we can approximate $(\lambda-c)/k\approx \lambda/k$. Applying the power-law ansatz $p(\lambda) \propto \lambda^{-\alpha_{\lambda}}$ to the master equation, and substituting $\alpha_{\lambda }=2-\gamma$, yields the transcendental equation:
\begin{equation}
a\gamma = 1 - k^{-\gamma}.
\label{eq:transend}
\end{equation}

Fig. 2e depicts the numerical solution for the intensity decay exponent $\alpha_\lambda=2-\gamma$ obtained from this transcendental equation as a function of $a$ and $k$. The dashed boundary corresponds to the $k = e^a$ stability threshold. 

Although the full stationary distribution $p(\lambda)$ is difficult to obtain, its moments can be analyzed analytically by examining the process in the event domain, specifically through the pre-event intensity $\lambda^-$. The distribution $p^-(\lambda^-)$ of the pre-event intensity is related to the stationary distribution $p(\lambda)$ via the Palm distribution formalism, where $p^-(\lambda^-) \propto \lambda^- p(\lambda^-)$. Working in the event domain, we can derive a recursive equation for the moments of $p^-(\lambda^-)$ (see Appendix~\ref{ap:theory}):
\begin{equation}
    \langle\left(\lambda^-\right)^n\rangle = \frac{1}{1+an-k^n}P_n\left(\langle\left(\lambda^-\right)^{n-1}\rangle, \dots\right),
    \label{eq:theory_stationary_lambda_}
\end{equation}
where $P_n(\dots)$ is a linear function of the lower-order moments of $\lambda^-$:
\begin{equation}
   P_n = \sum_{m=0}^{n-1}\binom{n}{m}\langle(\lambda^-)^{m}\rangle k^m c^{n-m}.
\end{equation}
The expectation $\langle \cdot \rangle$ is taken over the pre-event intensity distribution $p^-(\lambda^-)$. The first few moments are:
\begin{equation}
\begin{split}
    \langle\lambda^-\rangle &= \frac{c}{1+a-k},\\
    \langle\left(\lambda^-\right)^2\rangle &= \frac{c^2}{1+2a-k^2}\left(1+\frac{2k}{1+a-k}\right),\\
    \langle\left(\lambda^-\right)^3\rangle &= \frac{c^3}{1+3a-k^3}(...).
\end{split}
\label{eq:moments}
\end{equation}
The $n$-th moment diverges when the denominator is zero or negative, i.e., for $k^n \ge 1+na$. This divergence of moments is a direct reflection of the power-law tail in $p(\lambda)$ for $k > 1$.

The moments of the pre-event intensity can also be related to the moments of the stationary intensity via the equation:
\begin{equation}
    \langle f(\lambda^-)\rangle = \frac{\langle \lambda f(\lambda)\rangle}{\langle\lambda\rangle},
\end{equation}
where the averaging on the right-hand side is over $p(\lambda)$. This implies a simple relation for the $n$-th moment: $\langle\lambda^n\rangle = \langle\lambda\rangle \langle(\lambda^-)^{n-1}\rangle$. Thus, if the mean intensity $\langle\lambda\rangle$ is known, all higher moments of the distribution $p(\lambda)$ can be found analytically using the recursive formula for $\langle(\lambda^-)^{n-1}\rangle$.

When comparing finite-length simulations to the predicted stationary distribution, one should be cautious. As seen in Fig.~2f and 2g, while the bulk of the empirical distributions follows the predicted scalings, the tails can show significant deviations. These deviations are likely due to temporal correlations in $\lambda^-$, which can amplify stochastic fluctuations beyond what is expected from independent sampling.

\paragraph{Mean density}
Figure~2h shows numerical calculations for the mean event density $\rho$ as a function of the model parameters $a$ and $k$. While a general closed-form expression is elusive, analytical results can be obtained for four limiting cases:
\begin{enumerate}
    \item $k = 0$, constant reset. This case recovers the constant reset model, for which the mean density is $\rho = c(1-a)$, as given in Eq.~\eqref{eq:mean}.
    \item $a = 0$, no intensity decay. In this case, the intensity only changes at events according to the map $\lambda_{i+1} = k\lambda_i + c$. The stationary point of this map gives the mean intensity, yielding a density of $\rho = c/(1-k)$.
    \item $k = e^a$, stability boundary. At this boundary, the tail exponent of the intensity distribution is $\gamma=0$, which means $p(\lambda) \propto \lambda^{-2}$. The integral for the mean intensity, $\langle\lambda\rangle$, diverges logarithmically, causing the mean density to become infinite.
    \item $a = 1$, zero mean density. At this boundary, the mean inter-event interval (IEI) diverges (as in the constant reset model), and consequently, the mean event density becomes zero.
\end{enumerate}

A simple functional form that satisfies all the above four boundary conditions is:
\begin{equation}
    \rho(a, k) \approx c\frac{1-a}{1-ke^{-a}}.
    \label{eq:density_theoretical}
\end{equation}
A comparison of this phenomenological approximation to the simulated density is presented in Appendix~\ref{ap:mean_density}, showing a reasonably good fit across the parameter space.

\paragraph{Correlational structure}
The model's correlational structure can be analyzed from two distinct perspectives: dependencies in the continuous time domain, and correlations between discrete inter-event intervals in the event domain.
\begin{enumerate}
\item \textbf{Conditional intensity in the time domain.} We first examine the conditional intensity function, $h(\Delta t) = p(\text{event at } t+\Delta t|\text{event at } t)$. Figure~2i shows $h(\Delta t)$ for $a=1$ and various values of $k$. For large time lags ($\Delta t$), the function's decay appears to be universal, differing only by a scaling factor across different $k$. In contrast, the behavior at small $\Delta t$ is strongly controlled by $k$. Notably, for $k=2$, the conditional intensity diverges as $\Delta t \to 0$. We provide numerical support for these behaviors in Appendix~\ref{ap:condition_prob_tail} and Fig.~\ref{fig:autocorr_tail}.
    \item \textbf{Autocorrelation in the event domain.}
    We next analyze the correlation between the durations of IEIs. Due to the heavy-tailed nature of the IEI distributions, we use covariance of logarithms, $K_{\mathrm{ln}}(j) = \mathrm{cov}(\ln(\tau_{i +j}), \ln(\tau_i))$, which is more robust than Pearson's correlation. 
Figure~2j shows the correlation coefficient between neighboring events ($j = 1$) as a function of $a$ and $k$, calculated using simulations. For $ k > 0$, neighboring intervals are positively correlated and this correlation strengthens as $k$ increases. Conversely, for $k < 0$ neighboring intervals are negatively correlated. Figure~2k shows that these correlations decay exponentially with the separation $j$ in events, with a slower decay for larger $k$. 
\end{enumerate}

The above two perspectives reveals an interesting property in the model's structure. The process exhibits long-tailed (power-law) correlations in the time domain, which are driven by the heavy-tailed IEI distribution itself. However, when viewed simply as consecutive events (disregarding absolute time) it shows short-tailed i.e. exponentially decaying correlations in the event domain, which is commonly expected for Markov processes. 

To provide an analytical explanation for the exponential decay in the event domain, we can derive the auto-covariance function for the pre-event intensities.  As shown in Appendix~\ref{ap:theory_autocorr}, the covariance between pre-event intensities separated by $j$ events is:
\begin{equation}
    \operatorname{cov}(\lambda^-_{i+j}, \lambda^-_i) = \operatorname{var}(\lambda^-) \left(\frac{k}{a+1}\right)^j.
    \label{eq:autocorr_theory}
\end{equation} 
As follows from this expression, the correlations decay exponentially, with a factor $r=k/(a+1)$ at each step. There is, however, an important subtlety: the decay factor $r$ becomes greater than 1 for $k > a +1$, which implies that the covariance grows exponentially with the lag $j$. More critically, the covariance becomes ill-defined even earlier, as the variance of the pre-event intensity $\operatorname{var}(\lambda^-)$ diverges at $k^2 \ge 1+2a$, as follows from the equation for moments \eqref{eq:moments}. This indicates that the standard covariance is not a robust metric across the entire parameter space where the process itself is stable (up to $k < e^a$). In contrast, covariance of logarithms is well-defined across the whole range of parameters for stable systems which motivates it as the main metric for characterizing correlations between inter-event intervals. 

\subsection{Scale invariant model}
\label{sec:scale_invariant_model}
In this section, we focus on a special case of the linear jump scaling model (\ref{eq:general}) that has $c = 0$:
\begin{equation}
    d \lambda = -a \lambda^2 dt + (k-1) \lambda \eta
\label{eq:time_inv0}
\end{equation}
This equation corresponds to a high-intensity limit of the more general model(\ref{eq:general}),  where $(k-1)\lambda\gg c$ , and is particularly interesting because it remains invariant with respect to a rescaling of time. Both $a$ and $k$ are dimensionless, so only the initial value $\lambda(t = 0)$ introduces a scale for the process. 

As we shall see, the behavior of this equation is deceptive and requires careful treatment. Consider the ensemble average of both left- and right-hand sides:
\begin{equation}
    \langle d \lambda \rangle = \langle -a \lambda^2 dt + (k-1) \lambda \eta \rangle.
\label{eq:time_inv_average}
\end{equation}
After averaging over the Bernoulli-distributed event variable $\eta \sim \text{Ber}(\lambda dt)$, $\langle \eta \rangle = \lambda dt$, we obtain 
\begin{equation}
    d\langle \lambda \rangle =  (k-1-a) \langle\lambda^2\rangle dt
\label{eq:time_inv_average2}
\end{equation}

\begin{figure}
\centering
\includegraphics[width=0.3\linewidth]{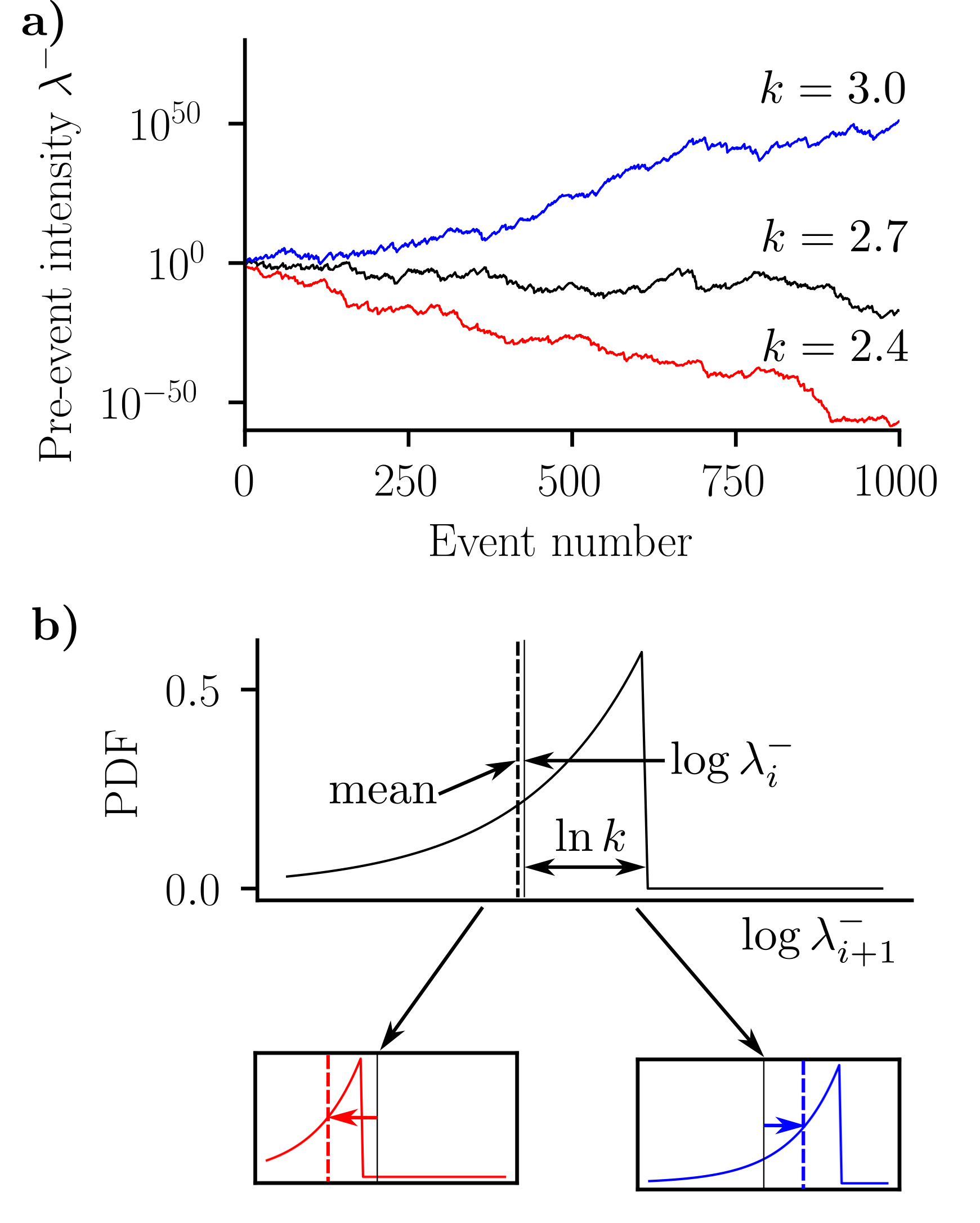}
\caption{\label{fig:scale_invariant} Behavior of the scale free-process near the critical regime $k=e^a$. 
a) The evolution of the pre-event intensity, $\lambda^-$, as a function of the event number for three values of $k$: subcritical ($k=2.4$), near-critical ($k=2.7$), and supercritical ($k=3.0$). For $k < e^a$, the process tends to decay, while for $k > e^a$, it tends to grow.
b) A schematic illustrating the biased random walk performed by the logarithmic intensity, $\ln(\lambda^-)$. The top panel shows the distribution of the step size, $\Delta(\ln\lambda^-)$, for each of the three regimes. The mean of this distribution (the drift) is negative for the subcritical case (red), positive for the supercritical case (blue), and near zero at the critical threshold (black), explaining the behaviors shown in (a).}
\label{fig:3}
\end{figure}

At first glance,  for $k > 1 + a$ it appears the mean value $\langle\lambda\rangle$ should diverge, because $\langle \lambda^2\rangle > \langle \lambda\rangle^2 > 0$ implies that the mean intensity grows perpetually. Counterintuitively, simulations such as in Fig.~3a reveal that  the actual critical threshold is $k= e^a$, and not $1+a$. Indeed, for $k < e^a$ the intensity almost always decays, even for $ k > 1+a$ (Fig. \ref{fig:3}a).

The apparent discrepancy between the divergent ensemble average $\langle\lambda\rangle$ and the decay observed in typical individual realizations for $k \in \left[1+a, e^a\right)$ can be resolved within the event-domain framework. By taking the logarithm of both parts of Eq. (\ref{eq:kesten_recursion__}) and assuming $c = 0$, the dynamics of the pre-event intensity, $\lambda^-$, are transformed into a simple random walk for $\ln\lambda^-$:
\begin{equation}
    \ln \lambda^-_{i+1} = \ln \lambda^-_{i} + \left(\ln k + a\ln U_i\right), \quad U_i \stackrel{\text{i.i.d.}}{\sim}\mathrm{Unif}(0,1).
\end{equation}
The step size, $\xi_i = \ln k + a\ln U_i$ is independent of the current process state, $\lambda^-_i$. The large-scale behavior of this random walk is governed entirely by its mean drift, $\langle\xi_i\rangle = \ln k - a$. This immediately establishes a critical threshold at $k_{\text{cr}} = e^a$, where the drift is zero. For $k>e^a$ the drift is positive, causing intensity to diverge to infinity, $\lambda^-_i \rightarrow \infty$. For $k<e^a$ the drift is negative, and thus every single trajectory almost surely decays to zero as the number of events approaches infinity.

This result sharpens the paradox: if every trajectory eventually decays for $k \in \left[1+a, e^a\right)$, how can their average $\langle\lambda\rangle$ diverge ? The resolution lies in the distinction between the behavior of a typical path and the properties of the entire ensemble's distribution at any given step. Despite the negative drift, at any given moment there exists a small and diminishing proportion of trajectories that, by chance, have experienced an unlikely sequence of positive steps. These trajectories can transiently achieve extremely high intensity values. Because the mean $\langle\lambda\rangle$ is calculated over the linear-scale intensity, these rare but enormous values dominate the ensemble average, causing it to diverge. It is not that some trajectories escape to infinity, but that the average is perpetually skewed by a transient, ever-changing population of outliers (e.g., as in Geometric Brownian Motion \cite{Oksendal2003}).

In the general reset model the additive term $c > 0$ acts as a floor, preventing the decay to zero, and ensuring stability for all $k < e^a$.

\subsection{Non-linear models}
\begin{figure}[h!]
\centering
\includegraphics[width=0.6\linewidth]{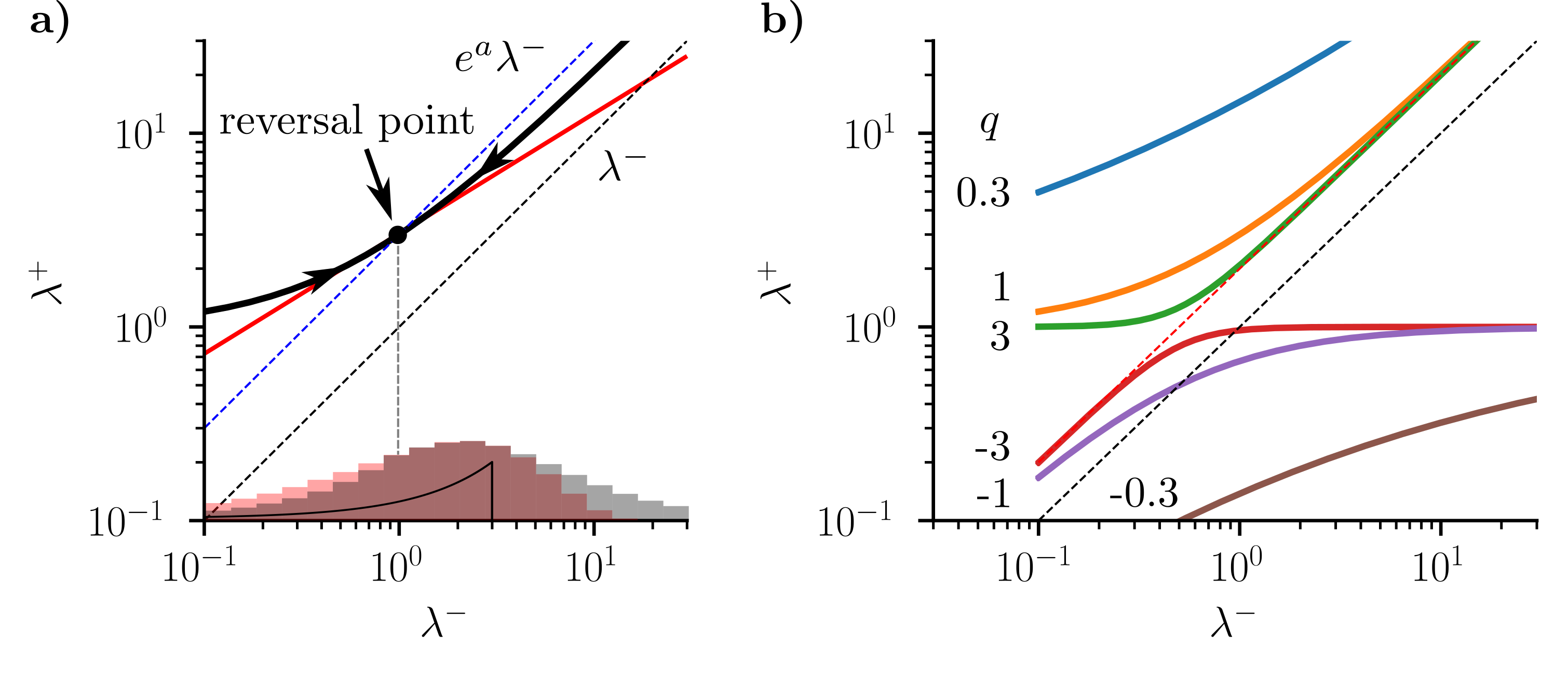}
\caption{Dynamics and forms of non-linear reset functions.
a) Dynamics of a process governed by a general non-linear reset function, $f(\lambda^-)$ (solid black curve), shown in the log-log plane. The intersection of the reset function with the critical line $\lambda^+ = e^a \lambda^-$ (dashed line) defines a reversal point. The process experiences a net drift towards this point, causing the stationary distribution of the pre-event intensity, $p(\lambda^-)$, to concentrate around it (gray histogram). This distribution is well-approximated by the corresponding canonical process (red line, red histogram). 
b) A family of analytically tractable reset functions, $f(\lambda)=(k\lambda^q+c)^{1/q}$, for several values of the non-linearity parameter $q$. The parameter $k$ is set to $2^q$ for illustration.
\label{fig:non_linear_processes}}
\end{figure}

We now extend our analysis to the general case of processes with a non-linear reset function, $\lambda^+=f(\lambda^-)$. While a full treatment is beyond the scope of this paper, we present an intuitive framework for understanding the dynamics of such processes, building on insights from the scale-invariant model. The key to understanding the system's behavior is the critical line, $\lambda^+=e^a\lambda^-$, which delineates regimes of positive and negative drift for the logarithmic intensity, $\ln \lambda^-$.  Specifically, if $f(\lambda^-)<e^a\lambda^-$, the drift is negative and pushes the intensity towards smaller values, and if $f(\lambda^-)>e^a\lambda^-$, the drift is positive and pushes the intensity towards larger values. Consequently, any intersection of the reset function with the critical line acts as a fixed point for the mean dynamics. We term such an intersection a \textit{reversal point}, as the sign of the average logarithmic drift reverses there. For a process with a single, stable reversal point, we hypothesize that the stationary probability density will concentrate in its vicinity. This behavior is demonstrated empirically in Figure \ref{fig:non_linear_processes}a (gray histogram) for an example non-linear process. Notably, the mode of the empirical distribution is slightly shifted to the right of the reversal point. This shift is likely due to the asymmetry of the stochastic decay from $\lambda^+$ to $\lambda^-$ (the PDF of the decay is shown as a black curve at the bottom of Fig. \ref{fig:non_linear_processes}a).

The above concentration of probability density near the reversal point suggests a powerful simplification. If the reset function $f(\lambda^-)$ is sufficiently smooth in the vicinity of the reversal point, its dynamics can be approximated by a simpler process that is linear in the log-log scale (red line in Fig. \ref{fig:non_linear_processes}). Such a process corresponds to the reset function
\begin{equation}
    f(\lambda) = p \lambda^q,    \label{eq:canonical_reset_function}
\end{equation}
where $p$ and $q$ are constants. We term such processes \textit{canonical}. Any non-linear reset function with a single reversal point can be locally approximated by such a canonical process, defined by matching the value and slope in the log-log plane at the reversal point. Therefore, the study of these canonical processes provides a general framework for understanding the bulk properties of a wide class of non-linear models.

The dynamics of a canonical process can be analyzed analytically. Taking the logarithm of the reset function \eqref{eq:canonical_reset_function} gives:
\begin{equation}
    \ln\lambda^+_{i}=q\ln\lambda^-_{i}+\ln p.
\end{equation}
By substituting the stochastic decay step, $\lambda^-_{i+1} = U_i^a\lambda^+_i$ yields a recursive equation for the pre-event intensity $\lambda^-$:
\begin{equation}
    \ln\lambda^-_{i+1}=q\ln\lambda^-_{i}+(a\ln U_i+\ln p).
\end{equation}
This is a first-order autoregressive (AR(1)) process for the logarithmic intensity, $y = \ln\lambda^-$:
\begin{equation}
    y_{i+1}=q y_{i}+\xi_i
\end{equation}
where $\xi = (a\ln U+\ln p)$ is an i.i.d variable. This equation is stable for $|q|<1$, and most its properties are analytically tractable. For example, the stationary mean of the logarithmic intensity is
\begin{equation}
    \langle y\rangle = \frac{\langle\xi\rangle}{1-q} = \frac{\ln p - a}{1 -q},
\end{equation}
Higher-order moments can be derived as well. The variance is
\begin{equation}
    \mathrm{var} (y) = \frac{\mathrm{var}(\xi)}{1-q^2} = \frac{a^2}{1 -q^2},
\end{equation}
The covariance function
\begin{equation}
    \mathrm{cov}(y_{i+j},y_i) = \mathrm{var}(y) q^j,
\end{equation}
indicates that correlations in the event domain decay exponentially. Unlike the linear reset model, the stability condition $|q|< 1$ required for recurrence also guarantees the convergence of the covariance function, thereby avoiding complications with diverging growth.

Beyond the linear and canonical models, another class of analytically tractable non-linear processes warrants mention, though a full analysis is reserved for future work. These are processes governed by the reset rule:
\begin{equation}
    f(\lambda) = (k\lambda^q +c)^{\frac{1}{q}},
\end{equation}
where $q \in \mathbb{R}$ is a parameter that controls the form of the non-linearity. As illustrated in Figure \ref{fig:non_linear_processes}b, the behavior of this function differs qualitatively for positive and negative values of $q$.  Models with $q>0$ are bounded from below and unbounded from above, allowing the process to generate bursts of arbitrarily high intensity, and thus very short IEIs. Conversely, models with $q<0$ are bounded from above and unbounded from below, allowing the process to generate bursts of very low intensities, and thus very long IEIs. This property is novel: while the standard linear reset model can produce individual long IEIs, the probability of observing several in succession is negligible as after a very long IEI, the intensity is reset back towards the baseline $c$.

One of the keys to analyzing this class of models is that the non-linear dynamics can be linearized through a variable transformation. By raising the reset function to the power of $q$, we can express the event-to-event update rule as
\begin{equation}
    \left(\lambda^+\right)^{q}=k\left(\lambda^-\right)^q+c.
    \label{eq:non_linear_update_1}
\end{equation}
This nonlinear recursion can be linearized using the variable transformation $y = \lambda^q$. Under this transformation, the reset rule becomes a simple linear update for the new variable:
\begin{equation}
    y^+ = k y ^-+c.
\end{equation}
Next, we transform the stochastic decay step. Recalling the event-domain rule $\lambda^-_{i+1}=\lambda^+_{i} U_i^a$, and raising both sides to the power of $q$, we get
\begin{equation}
    \left(\lambda^-_{i+1}\right)^q=\left(\lambda^+_{i}\right)^q U_i^{aq},
\end{equation}
which in terms of the new variable is simply
\begin{equation}
    y_{i+1}^+ = y_i^- U_i^{aq}
\end{equation}
These equations demonstrate that the dynamics of the transformed variable $y$ are governed by a process that is mathematically identical to the linear jump scaling model, but with an effective decay parameter of $a' = aq$. This mapping allows all the results derived for the linear model --- such as stability conditions, moment calculations, and tail exponents --- to be directly applied to this nonlinear case after the appropriate transformation. A detailed exploration of this model class is a promising direction for future work.

\section{Simulation and Fitting}
This section outlines some practical methods for simulating the model, estimating its parameters from data, and validating the model's goodness-of-fit.

\paragraph{Simulation}
The point processes described in this work can be simulated using two approaches. The first is a time-stepping method, evolving the intensity $\lambda(t)$ with a small time step $dt$ according to the stochastic differential equation (Eq.~\ref{eq:general}).

A more efficient method, however, is to simulate the process in the event domain, stepping directly from one event to the next. This event-stepping algorithm is as follows:
\begin{enumerate}
    \item Given the intensity just before the $i$-th event, $\lambda_i^-$, calculate the post-event intensity $\lambda_i^+ = f(\lambda_i^-)$ using the deterministic reset rule (e.g., Eq.~\eqref{eq:iterative}).
    \item Determine the next pre-event intensity, $\lambda_{i+1}^-$, according to Eq.~\eqref{eq:sampling_identity}.
    \item The inter-event interval $\tau_i$ can then be expressed through the intensity change:
\begin{equation}
    \tau_i = \frac{1}{a}\left(\frac{1}{\lambda^-_{i+1}}-\frac{1}{\lambda^+_{i}}\right).
\end{equation}  
\end{enumerate}
The above approach avoids computationally expensive integration during long periods of inactivity, and is therefore significantly faster especially when the distribution of IEIs has a long tail.

\paragraph{Parameter Estimation}
The model parameters $(\lambda_0, a, k, c)$ can be estimated from an observed sequence of event times $\{t_1, t_2, \dots, t_N\}$ using the method of Maximum Likelihood Estimation (MLE). The likelihood of the observed sequence is the product of the conditional probabilities of each inter-event interval, $\tau_i = t_{i+1} - t_i$. The log-likelihood function is therefore:
\begin{equation}
    \mathcal{L}(\theta) = \sum_{i=1}^{N-1} \ln p(\tau_i | \lambda_i^+(\theta)),
\end{equation}
where $\theta$ represents the set of model parameters and $\lambda_i^+$ is determined by the history of events and the parameters. The parameters can be found by numerically minimizing the negative of this log-likelihood.

Due to the heavy-tailed nature of the distributions, a large number of events is required for robust parameter estimation. Our empirical tests suggest that on the order of $N=10^4$ events are needed to achieve typical uncertainties of about 5\% for the key parameters ($a$, $k$, and $c$).

\paragraph{Model Validation}
Beyond fitting for model parameters, it is typically critical to be able to assess whether the model appropriate for the dataset at hand (i.e. address potential model mismatch). A simple check is to compare statistical properties, such as the empirical IEI tail exponent or event-domain correlations, with the model's predictions (e.g., Fig 2j, k).

A more systematic approach is to use the Probability Integral Transform (PIT) and analyze the resulting residuals. For a correctly specified model, the transformed variables $u_i = F(\tau_i | \lambda_i^+)$, where $F$ is the model's cumulative distribution function (CDF), should be uniformly distributed on $[0, 1]$. By testing these residuals for uniformity (e.g., using a Kolmogorov-Smirnov test), particularly when conditioned on the state of the process (such as the previous interval $\tau_{i-1}$), one can diagnose specific ways in which the model may fail to capture the data's temporal structure. An example of application of this approach is presented in Appendix~\ref{ap:validation} and Fig.~\ref{fig:s_validation}.

\begin{figure}
\centering
\includegraphics[width=1.\linewidth]{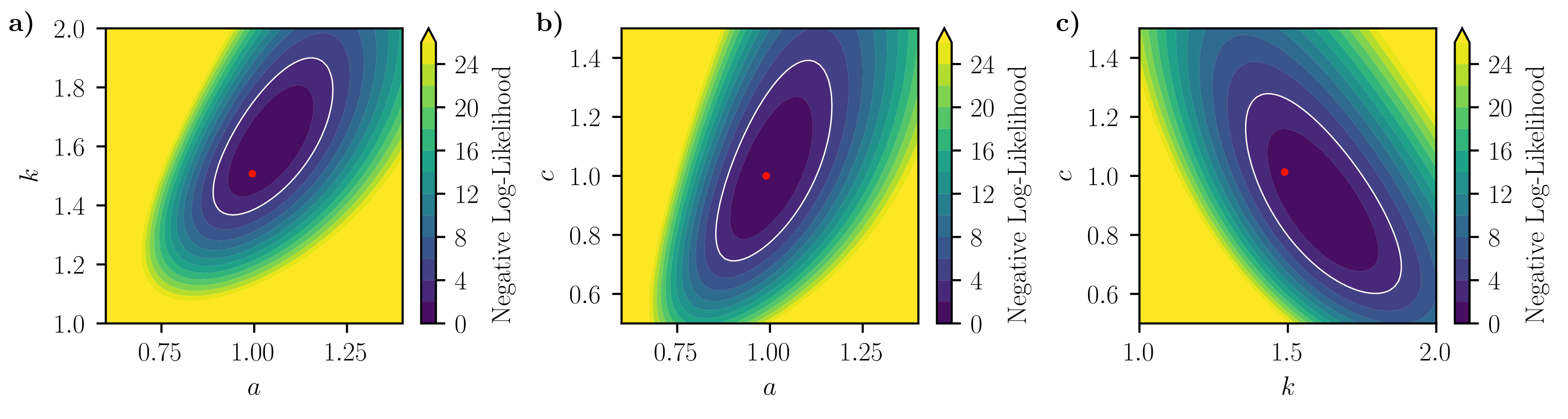}
\caption{\label{fig:s_validation} An example of the fitting of a model with $a = 1$ and $k = 1.5$ (indicated by the red dot) for $N_{\text{events}} = 10^3$. The plots show the negative log-likelihood of the model, as a function of the corresponding pair of parameters. The white line indicates $95\%$ MLE confidence estimate.
}
\label{fig:4}
\end{figure}

\section{Conclusion and Discussion}

In this work, we demonstrated that purely local intensity dynamics (without invoking global memory) can naturally generate power-law tails in the distributions of inter-event intervals (IEIs) and/or event rates. Specifically, we showed that it is sufficient to have quadratic decay of event intensity between events combined with a fairly general form of an intensity jump after events. Our model can generate all possible IEI exponents corresponding to normalizable distributions ($\alpha_{IEI} > 1$), and, in the case of the linear intensity jump function, all event rate exponents for distributions with finite mean ($\alpha_{\lambda} > 2$). 

Depending on the reset function, we showed that our model can yield positive or negative correlations between neighboring inter-event intervals. The correlational structure of events demonstrated another interesting feature: while the excess intensity (a rescaled version of autocorrelation function) always decays as a power law, the correlations between inter-event intervals decay exponentially as a function of the number of events between them.  

We also established an intuitive framework for understanding the behavior of the process with a general, non-linear reset function. In logarithmic coordinates the pre-event intensity $\lambda^-$ evolves as a biased random walk. The direction of the bias, or drift, is governed by whether the reset function lies above or below a critical line defined by $\lambda^+ = e^a \lambda^-$. This mechanism leads the stationary distribution of the process intensity to cluster around reversal points, where the reset function crosses this critical boundary.

In addition to analytically and numerically characterizing the model, we discussed methods for parameter estimation (using maximum likelihood) and model validation (via conditional Probability Integral Transform residuals). Due to the inherent power-law nature of the involved distributions, parameter estimates converge slowly, typically requiring around  $\sim 10^4$ events to achieve accuracy within about $5\%$. This consideration should inform experimental designs aimed at empirically testing models for data with long tails in general. 

We view our framework as an alternative to nonlinear Hawkes processes or Cox-like processes. While these frameworks are sufficiently rich to reproduce power-law distributions for inter-event intervals and/or intensity (we demonstrate this using examples in the Appendix \ref{ap:pl_hawkes} and \ref{ap:pl_cox}, Fig. \ref{fig:hawkes_cox_power_law}), each comes with characteristic limitations. In Cox-like processes, events have no influence on the intensity dynamics, limiting their applicability to natural and behavioral phenomena where event-driven feedback is expected. Nonlinear Hawkes processes, on the other hand, couple the size of rate increases after each event tightly to the model's overall nonlinear mapping from accumulated "tension" (a hidden variable tracking past influences) to the event rate; this prevents independent tuning of how the rate decays between events versus how much it jumps right after one. For the common case of exponential kernels, this coupling is explicit: the process can be reformulated as a local one where the decay dynamics rigidly determine the form of the reset function (see Appendix \ref{ap:Hawkes_markov_mapping}). In essence, these model classes either lack any connection between event occurrences and rate evolution (Cox-like) or impose complicated constraints on that connection (nonlinear Hawkes).
 
Our model avoids the above-mentioned deficiencies by providing independent control of the intensity decay dynamics and post-event resets. Such flexibility could be advantageous when modeling phenomena involving large bursts after long periods of inactivity, or generating specific correlation structures under some statistical constraints. Additionally, the locality of the dynamics in our model gives it a mechanistic appeal that is also generally lacking in nonlinear Hawkes processes, which rely on non-local kernels. This is the motivation behind why one might want to recast, say, nonlinear Hawkes processes with exponential kernels in local form.

We of course do not mean to propose that our model replaces Cox-like processes or nonlinear Hawkes processes, but rather that these are complementary model classes each with their own domains of applicability. Indeed, the local nature of our model may make it an inappropriate description for some phenomena that are thought to have explicit (complex) memory mechanisms, and nonlinear Hawkes processes would provide a more natural description. The burstiness in some phenomena may also be more appropriately conceptualized as originating from fluctuating external drives, which can be directly expressed in Cox-like models. Lastly, the inherent power-law distributions in much of the parameter space of our model could also make it not the best choice for phenomena in which long-tail distributions are not a prominent feature. Ultimately, selecting between these frameworks is a matter of mechanistic plausibility and practical convenience. 

The models presented here represent only a subset of the possibilities within our framework. While this paper focused on generating bursty dynamics with a positive decay parameter ($a>0$), the framework also accommodates a negative parameter ($a<0$). A negative decay parameter defines a process where intensity grows between events, yielding short-tailed (sub-exponential) IEI distributions and regular, rather than bursty, event patterns (See Appendix \ref{ap:negative_a}, Fig. \ref{fig:negative_a}). A complete systematization of these distinct dynamic regimes, explored across a full variety of reset functions, is an essential next step. Such an analysis is necessary for a comprehensive understanding of the framework's full theoretical scope and the complete range of behaviors it can generate.

Our framework could be extended in several important ways. Perhaps most prominently, the model could be extended to marked point processes, where each event (for example, a financial transaction) is associated with a magnitude (e.g., transaction size or volume). Such a generalization would be crucial for modeling not only financial markets, but also a variety of natural phenomena such as solar flares or earthquakes. Fully describing the latter systems could further require an explicit treatment of  spatial dynamics. 

Another promising application of our model is in modeling animal and human behaviors, where events represent actions such as grooming, engaging with a particular object, or social communications. In these applications, the quadratic intensity decay would govern changes in behavioral policies during periods without action execution, and thus capture statistical properties of inter-action intervals. Realistic modeling of behavioral dynamics would likely require extending our framework to multivariate point processes, for example to account for multiple possible types of actions or objects/others interacted with. Depending on the complexity of the experimental settings, the model might also need to explicitly incorporate environmental variables to account for the availability of activities. 

\paragraph{Acknowledgments.} We thank Michele Nardin for detailed comments on the manuscript's structure and clarity and Didier Sornette for valuable feedback on its context, motivation, and suggestions. We are also grateful to Virginia Ruetten, Tosif Ahamed, Aneesh PremaBalakrishnan, and Xiao Liu for insightful discussions of this work.


\section{Appendix}

\appendix

\section{Experimental and theoretical literature about long-tail distributions}

\begin{table}[h]
    \centering
    \begin{tabular}{|p{3.4cm}|p{1.9cm}|p{1.9cm}|p{5cm}|}
        \hline
        \textbf{Paper / Reference}   & \textbf{IEI} & \textbf{Freq.}&  \textbf{notes} \\
        \hline
        \multicolumn{4}{|c|}{Trades} \\    
        \hline
        Ivanov, 2004 \cite{common_scaling1} & Weibull & & Inter-trade times are weibul distributed with $\delta = 0.74$ \\
        \hline
        Plerou, 2000 \cite{plerou2000} & & $\sim 2$ decades  & power law for the number of transaction per unit of time, $\alpha_\lambda \approx 3.4$\\
        \hline
        \multicolumn{4}{|c|}{Forest fires} \\    
        \hline
        Song, 2006 \cite{SONG2006527} & $\sim 2$ decades  & & Forest fires may have a power law for inter-event time, $\alpha_{\text{IEI}}\approx 2.2$\\
        \hline
        \multicolumn{4}{|c|}{Earthquakes}  \\    
        \hline
        Ogata, 1988 \cite{ogata1988} & & $\sim 2$  decades, with exp. cutoff & Fitting earthquake data with $1/t$ Hawkes, which theoretically should give a power law for frequencies, empirically $\alpha_\lambda \approx 1$\\
        \hline
        Corral, 2004 \cite{corral_long-term_2004} & $\sim 2.5$ decades with exp. cutoff &   & Earthquake recurrence PDF is universal, power law with an exponential cutoff, $\alpha_{\text{IEI}} \approx 1$\\
        \hline
        \multicolumn{4}{|c|}{Solar flares}  \\    
        \hline
        Wetland, 2002 \cite{wheatland_understanding_2002} Baiesi, 2006 \cite{Baiesi2006} & $\sim 3$ decades &  & power law for waiting times for solar flares, $\alpha_{\text{IEI}} \approx 2.2$\\
        \hline
        \multicolumn{4}{|c|}{Communication}  \\    
        \hline
        Clauset, 2009 \cite{clauset2009} & & $2$ to $3$ decades  & possible power law for the number of calls per day, $\alpha \approx 2.1$. \\
        \hline
        Gandica, 2017 \cite{Gandica_2017} & 2.5-3 decades & & Power law for inter-tweet time on twitter ($\alpha_{\text{IEI}} \approx 1.1$) and Wikipedia posts ($\alpha_{\text{IEI}} \approx 1.1$)\\
        \hline
        Zhao et al, 2011 \cite{zhao_2011} &3 decades && Power law for inter-events for SMS messages, $\alpha_{\text{IEI}} \approx 1.2$ \\
        \hline
        \multicolumn{4}{|c|}{various human activity}  \\   
        \hline
        Hong, 2007 \cite{HONG20076} & & $\sim 1$ decade & power-law in the number of bankruptcies per day, $\alpha_{\lambda} \approx 0.9$\\
        \hline
        Vasques et al, 2006 \cite{Vazques2006} & 2.5 decades& & IEI power law for library loans, $\alpha_\lambda \approx 1$ \\
        \hline
        Henderson et al, 2001, \cite{henderson_modelling_2001} & 1.5 decades & & power law for inter-login times for users for networked games , $\alpha_{\text{IEI}} \approx 2.2$\\
        \hline
        Zhou et al. 2008 \cite{zhou_role_2008} & 2 decades & & IEI for rating movies is user-dependent power law with $\alpha_{\text{IEI}} \in [1.5, 2.5]$, with larger values for more active users \\
        \hline
        
    \end{tabular}
    \caption{Summary of empirical papers with bursty behavior. The decades reported in the IEIs and frequencies columns refer to the number of decades over which the power law holds.}
    \label{tab:bursty_models}
\end{table}

\begin{table}[h]
    \centering
    \begin{tabular}{|p{3.4cm}|p{1.4cm}|p{1.4cm}|p{2.1cm}|p{5cm}|}
        \hline
        \textbf{Paper / Reference}   & \textbf{IEI} & \textbf{Freq} & \textbf{locality} & \textbf{notes} \\
        \hline
        \multicolumn{5}{|c|}{ Priority Queuing} \\
        \hline
        Barabási (2005) \cite{barabasi_origin_2005}  & \ding{51} &  & \ding{51}  &  Tasks executed by priority cause some to be delayed, resulting in heavy-tailed waiting times.\\
        Vázquez et al. (2006) \cite{Vazques2006}& \ding{51} &  & \ding{51}  & \\
        \hline
        \multicolumn{5}{|c|}{Hawkes-like processes} \\
        \hline
        Jo et al. (2015) \cite{Jo2015}& \ding{51} &  & no & $1/t$ kernel with finite memory size\\
        Kanazawa (2023) \cite{kanazawa_asymptotic_2023} & & \ding{51} & no (except exp. kernels) & General Nonlinear Hawkes processes\\
        Kanazawa (2021) \cite{kanazawa_ubiquitous_2021}& & \ding{51} & no (except exp. kernels) & Nonlinear Hawkes, power law under special conditions\\
        \hline
        \multicolumn{5}{|c|}{Self-reinforcement Two-State Model } \\
        \hline
        Karsai et al. (2012)\cite{karsai_universal_2012}  & \ding{51} &  & no & Alternates between active and inactive states with reinforcement, producing clusters. \\
        \hline
        \multicolumn{5}{|c|}{Cox-like processes}\\
        \hline
        Malmgren et al. (2008) \cite{Malmgren2008}  & \ding{51} &   & \ding{51}  & Effective power law due to cycles \\
        Biró (2005) \cite{Biro2005} &  & \ding{51} & \ding{51} & multiplicative noise \\
        Kleinberg (2003) \cite{kleinberg_bursty_nodate} &  &  & \ding{51} & poisson automaton, burstiness \\
        \hline
        \multicolumn{5}{|c|}{Heavy-Tailed Renewal Process}\\
        \hline
        Ross (1996) \cite{ross1995stochastic} &\ding{51} & & & IEI is power-law by construction \\
        \hline
        \multicolumn{5}{|c|}{Threshold/Cascade Models (SOC)} \\
        \hline
        Bak (1988) \cite{Bak1998} &  &  \ding{51} & \ding{51} &\\
        \hline
        Kesten, 1973\cite{kesten1973} &  & \ding{51} & \ding{51} & Random process with power law for the intensity\\
        \hline
        \multicolumn{5}{|c|}{Our model}\\
        \hline
        Current paper & \ding{51} & \ding{51} & \ding{51} & \\
        \hline
    \end{tabular}
    \caption{Summary of models for bursty behavior in social dynamics.}
    \label{tab:bursty_models}
\end{table}

\section{Scaling}
\label{ap:scaling}

We now examine how the dynamical equation (\ref{eq:general}) transforms under a rescaling of time, $t' = s t$. Under this transformation, the intensity $\lambda$, defined as the event generation probability per unit time ($dp/dt$), rescales as follows:

\begin{equation}
\lambda' = \frac{dp}{dt'} = \frac{\lambda}{s}.
\end{equation}

Substituting this relation into the stochastic equation for the linear reset model (\ref{eq:general}) yields:

\begin{equation}
s d\lambda' = -a s^2 (\lambda')^2 s^{-1} dt' + (k s \lambda' + c) \eta.
\end{equation}

Simplifying by dividing by $s$, we obtain:

\begin{equation}
d\lambda' = -a (\lambda')^2 dt' + \left(k\lambda' + \frac{c}{s}\right)\eta.
\end{equation}

This demonstrates that models differing solely in the parameter $c$ (while sharing the same values of $a$ and $k$) are simply time-rescaled versions of one another. Consequently, the event density and the Palm function scale proportionally with $c$.

\section{Master equation derivation}
\label{ap:master}

Here we derive the master equation (\ref{eq:fokker2}) starting from the stochastic equation (\ref{eq:general}), obtaining equation (\ref{eq:fokker0}) as a special case. Recall the stochastic process for the event intensity $\lambda(t)$ governed by:
\begin{equation}
    \mathrm{d}\lambda = -a\lambda^2\,\mathrm{d}t + [(k-1)\lambda + c]\,\eta,
\end{equation}
where $a$, $k$, and $c$ are constants, and $\eta$ is an event indicator variable, equal to $1$ if an event occurs during $[t, t+\mathrm{d}t)$ and $0$ otherwise. The probability of an event occurring is thus given by $P(\eta = 1) = \lambda\, \mathrm{d}t$.

Let $P(\lambda, t)$ denote the probability density of $\lambda$ at time $t$. The probability balance over a small interval $\Delta t$ gives:
\begin{equation}
    P(\lambda, t+\Delta t) = \int_0^\infty \mathrm{d}\lambda'P(\lambda', t)\sum_{\eta\in\{0,1\}} P(\eta|\lambda')\,\delta\left(\lambda - [\lambda' - a\lambda'^2\Delta t+((k-1)\lambda'+c)\eta]\right).
\end{equation}

Summation over $\eta$ using $P(1|\lambda')=\lambda'\Delta t$ and $P(0|\lambda')=1-\lambda'\Delta t$ yields:
\begin{align}
    P(\lambda, t+\Delta t) = \int_0^\infty \mathrm{d}\lambda' P(\lambda', t)\Big\{&\lambda'\Delta t\,\delta\left(\lambda - [k\lambda' + c ]\right) \nonumber\\
    &+(1 - \lambda'\Delta t)\,\delta\left(\lambda - [\lambda' - a\lambda'^2\Delta t]\right)\Big\} + o(\Delta t).
\end{align}

Expanding the delta functions to first order in $\Delta t$:
\begin{equation}
    \delta\bigl(\lambda - [\lambda' - a\lambda'^2\Delta t]\bigr)
    = \delta(\lambda - \lambda') + a\lambda'^2\Delta t\,\partial_{\lambda}\delta(\lambda - \lambda') + o(\Delta t).
\end{equation}

Substituting this expansion into the integral and performing integration, we obtain:
\begin{align}
    P(\lambda, t + \Delta t) =& P(\lambda, t)\\ &+\Delta t \left[\frac{\partial}{\partial\lambda}\left(a\lambda^2 P(\lambda, t)\right) - \lambda P(\lambda, t) + \int_0^\infty \mathrm{d}\lambda' \lambda' P(\lambda', t)\delta(\lambda - [k\lambda'+c])\right] + o(\Delta t).
\end{align}

Evaluating the integral via a change of variables $\lambda' = (\lambda - c)/k$, with Jacobian $1/k$, we get:
\begin{equation}
    \int_0^\infty \mathrm{d}\lambda' \lambda' P(\lambda', t)\,\delta(\lambda - [k\lambda'+c]) = \frac{(\lambda - c)}{k^2}\,P\left(\frac{\lambda - c}{k}, t\right).
\end{equation}

Taking the limit $\Delta t \to 0$, we arrive at the master equation (\ref{eq:fokker2}):
\begin{equation}
    \frac{\partial P(\lambda, t)}{\partial t} = \frac{\partial}{\partial \lambda}\left(a\lambda^2 P(\lambda, t)\right) - \lambda P(\lambda, t) + \frac{(\lambda - c)}{k^2}P\left(\frac{\lambda - c}{k}, t\right).
\end{equation}

Here, the first term describes deterministic drift due to quadratic decay. The second and third terms correspond to intensity jumps: the second term accounts for probability flowing out of the state $\lambda$ due to jumps triggered by events, while the third term describes probability influx from states that jump into $\lambda$.

For $\lambda < c$, the last term vanishes, as $P(\lambda)=0$ for $\lambda < 0$, simplifying the equation to:
\begin{equation}
    \frac{\partial P(\lambda, t)}{\partial t} = \frac{\partial}{\partial \lambda}\left(a\lambda^2 P(\lambda, t)\right) - \lambda P(\lambda, t),
\end{equation}
Assuming the time derivative to be zero, we
recover equation (\ref{eq:fokker0}), the master equation for constant reset processes with $\lambda < \lambda_0 \equiv c$.

\section{Fractal structure analysis}
\label{ap:fractal}
To understand the scaling of empty and black regions, let us assume that at particular scale it is impossible to resolve gaps that are smaller than $\tau$. As we zoom out, $\tau$ is going to increase, and some gaps are going to disappear. Assuming the distribution of the existing gaps to be described by the survival function $S(\tau)$ with a cutoff at $\tau$, the proportion of gaps that are going to close is going to be
\begin{equation}
    P(\tau)d\tau = -\frac{S'(\tau)d\tau}{S(\tau)}  = - (\ln S(\tau))'d\tau
    \label{eq:gaps1}
\end{equation}

The mean duration of the regions with events, $l$, is going to increase by the size of the gaps that disappeared plus the size of the regions that have been attached:
\begin{equation}
\begin{split}
    dl &= \tau P(\tau)d\tau + l P(\tau)d\tau 
\end{split}    
\end{equation}
After using the expression (\ref{eq:gaps1}), one can solve the resulting equation to get the following solution:
\begin{equation}
    l(\tau) = S^{-1} \int_0^\tau S(t) dt - \tau
\end{equation}
Assuming the equation for PDF (\ref{eq:inter_event1}), and without a loss of generality taking $\lambda_0 = 1$, we obtain:
\begin{equation}
    l(\tau) = \frac{1 + \tau - (1+a\tau)^{1/a}}{a-1}
\end{equation}
for $a\neq 1$, while for $a = 1$, the expression for the mean duration is the following
\begin{equation}
    l(\tau) = (1+\tau) \ln ({1+\tau}) - \tau
\end{equation}

One can notice a significant difference in the behavior of $l/\tau$ for $a > 1$ and $a < 1$ in the limit $\tau\rightarrow\infty$. While for $a>1$ $l/\tau$ goes to a constant value $(a-1)^{-1}$ (i.e. the proportion of active vs empty regions stays constant), in the case of $a<1$ $l/\tau$ increases to infinity, which implies that the whole event plot becomes black.

For the interested reader, the mean number of events in a block is easy to obtain, by noticing that the number of blocks is equal to total number open gaps (up to 1), i.e.
\begin{equation}
    N_{\text{blocks}} = N_{\text{events}} S(\tau), 
\end{equation}
which makes the mean number of events per block
\begin{equation}
    n(\tau) = S^{-1}(\tau).
\end{equation}
The mean duration of the blocks relative to the mean number of points in a block, $l(\tau)/n(\tau)$, has the following form (for $a\neq 1$):
\begin{equation}
    \frac{l(\tau)}{n(\tau)} = \frac{-1 + (1 + \tau) (1 + a \tau)^{-1/a}}{a-1},
\end{equation} 
And either goes to a constant if $a<1$, or infinity if $a>1$ (and goes to infinity logarithmically if $a = 1$).

Notice that in the whole reasoning about the mean duration we did not rely on any correlational structure between events. Where the correlations start playing a role is in the \textit{distribution} of the durations of the blocks, which could be measured either in terms of time, or in terms of the number of events within a block. With respect to the distribution of the number of events  within a block, since in our particular case each next interval between events is independent from the previous one, each inter-event interval has the same probability to be larger than $\tau$. This behavior makes an exponential distribution of the number of events in a block.

\section{Probability distribution for inter-event intervals for the general model}
\label{ap:iei_prob_distrib}
In the main text we claimed that for the general model (\ref{eq:general}), the inter-event interval distribution scales as $(a\tau)^{-1-a/1}$, which we intend to prove in this section. 
The full expression for the inter-event interval distribution can obtained by the direct substitution of the equation (\ref{eq:inter_event2}) into (\ref{eq:inter_event2}):
\begin{equation}
p(\tau) = \int_{0}^{\infty} \lambda^+\left(1 + \lambda^+a\tau\right)^{-1/a - 1}p(\lambda^+)d\lambda^+.
\end{equation}
To prove that the quantity behaves as $\tau^{-1/a-1}$, we divide the expression by $\tau^{-1/a-1}$:
\begin{equation}
p(\tau)/\tau^{-1/a-1} = \int_{0}^{\infty} \lambda^+\left(1/\tau + \lambda^+a\right)^{-1/a - 1}p(\lambda^+)d\lambda^+.
\end{equation}
In the limit $\tau \to \infty$, if $p(\lambda^+)$ is non-zero only for $\lambda^+>\lambda_{\min}$, one can neglect the term $1/\tau$, as the equation becomes:
\begin{equation}
\begin{split}
p(\tau)/\tau^{-1/a-1} & \to \int_{\lambda_{\min}}^{\infty}  \lambda^+\left( \lambda^+a\right)^{-1/a - 1}p(\lambda^+)d\lambda^+ \\
& = a^{-1/a-1}\int_{\lambda_{\min}}^{\infty} \left( \lambda^+\right)^{-1/a}p(\lambda^+)d\lambda^+\\
\end{split}
\end{equation}
the last expression is a finite constant if the integral converges, which can be easily seen, since
\begin{equation}
        \int_{\lambda_{\min}}^{\infty} \left( \lambda^+\right)^{-1/a}p(\lambda^+)d\lambda^+<\lambda_{\min}^{-\frac{1}{a}}\int_{\lambda_{\min}}^{\infty}p(\lambda^+)d\lambda^+= \lambda_{\min}^{-\frac{1}{a}}.
\end{equation}
 
The value of $\lambda_{\min}$ can be found from the iterative equation $\lambda^+ = k\lambda^-+c$:
\begin{itemize}
    \item $k>0$. Here $\lambda_{\min} = c$, which corresponds to $\lambda^- = 0$.
    \item $k<0$. Here $\lambda_{\min} = c(1+k)$, which corresponds to $\lambda^- = c$ (the largest possible value)
\end{itemize}

\section{Stability analysis}
\label{ap:stability_analysis}
Here we analyze the stability of the linear jump scaling model \eqref{eq:general}. The analysis relies on the mapping of our model in the event domain onto a Kesten process \cite{kesten1973}:
\begin{equation}
    \lambda^+_{i+1} = k U_i^a \lambda^+_i + c, 
\end{equation}
where $U_i \stackrel{\text{i.i.d.}}{\sim}\mathrm{Unif}(0,1)$. A Kesten process of the form $X{n+1} = A_n X_n + B_n$ is positive recurrent if $\langle \ln {A_i}\rangle <0$. In our case $A_i = k U_i^a$, and the stability condition is therefore
\begin{equation}
\begin{split}
\langle \ln |k U_i^a| \rangle &< 0 \\
\ln |k| + a \langle \ln U_i \rangle &< 0\\
\ln |k| - a &< 0,
\end{split}
\end{equation}
which yields the stability condition $ |k| < e^a$. For $k \ge -1$ and $a > 0$ this simplifies to $ k < e^a$.

\section{Derivation of the Asymptotic Tail Exponent}

\label{ap:transcendental_derivation}

To find the asymptotic behavior of the stationary distribution $p(\lambda)$, we start with the full master equation (Eq.~\ref{eq:fokker2}):
\begin{equation}
    \frac{\partial}{\partial\lambda}\left(a\lambda^2p(\lambda)\right) - \lambda p(\lambda) + \frac{1}{k^2} p\left(\frac{\lambda - c}{k}\right)(\lambda - c) = 0.
\end{equation}
We substitute the power-law ansatz $p(\lambda) \propto \lambda^{\gamma - 2}$ directly into this equation. The terms become:
\begin{itemize}
    \item $\frac{\partial}{\partial\lambda}\left(a\lambda^2 \cdot \lambda^{\gamma-2}\right) = a\gamma\lambda^{\gamma-1}$
    \item $\lambda p(\lambda) \propto \lambda^{\gamma-1}$
    \item $\frac{1}{k^2} p\left(\frac{\lambda-c}{k}\right)(\lambda-c) \propto \frac{1}{k^2}\left(\frac{\lambda-c}{k}\right)^{\gamma-2}(\lambda-c) \propto \frac{\lambda^{\gamma-1}}{k^\gamma}$
\end{itemize}
Combining these gives the expression:
\begin{equation}
    a\gamma\lambda^{\gamma-1} - \lambda^{\gamma-1} + \frac{\lambda^{\gamma-1}}{k^\gamma} \approx 0.
\end{equation}
Dividing by the common factor $\lambda^{\gamma-1}$ yields $a\gamma - 1 + k^{-\gamma} = 0$, which rearranges to the final transcendental equation for the exponent $\gamma$:
\begin{equation}
    a\gamma = 1 - k^{-\gamma},
\end{equation}
which matches Eq.~\eqref{eq:transend} in the main text. 

\section{Moments of the pre-event intensity probability distribution}
\label{ap:theory}
The equation (\ref{eq:iterative_prob_distrib}) allows for an analytical derivation of the dynamic equation for the moments of the probability distribution. 
One derives for the n-th moment:
\begin{equation}
\begin{split}
    \langle (\lambda_{i+1}^-)^n \rangle & = \int_0^{\lambda^+_{i}}\lambda^n  p(\lambda|\lambda^+_i)d\lambda\\ 
    &= \int_0^{\lambda^+_{i}}\lambda^n  \frac{1}{a\lambda}\left(\frac{\lambda}{\lambda^+_i}\right)^{1/a}d\lambda\\
    & = \frac{(\lambda^+_{i})^n}{1+an}.
\end{split}
\end{equation}

After expressing $\lambda^+_{i} = k\lambda^-_i+c$, and taking the average, we obtain a recursive formula:

\begin{equation}
    \langle (\lambda^-_{i+1})^n \rangle = \frac{\langle(k\lambda^-_i+c)^n\rangle}{1+an}.
    \label{eq:rec1}
\end{equation}

After expanding the linear expression in the parenthesis, we obtain the collection of the dynamic equations:
\begin{equation}
    \langle \lambda_{i+1}^n \rangle = \frac{k^n}{1+an}\langle\lambda_i^n\rangle +\frac{P_n(\langle\lambda_{i}^{n-1}\rangle,...)}{1+an},
    \label{eq:rec2}
\end{equation}
where $P_n(\cdot)$ is a linear function of the lower moments:
\begin{equation}
   P_n(\langle\lambda_{i}^{n-1}\rangle,...) = \sum_{m=0}^{n-1}\tbinom{n}{m}\langle\lambda^{m}\rangle k^m c^{n-m}.
\end{equation}

One  can then solve equation (\ref{eq:rec2}) to find a stationary solution (\ref{eq:theory_stationary_lambda_}).

\section{Autocorrelation function for the pre-event intensity}
\label{ap:theory_autocorr}
Here we derive the auto-covariance function (\ref{eq:autocorr_theory}) for the pre-event intensities.
Consider the dynamic equation for $\langle\lambda_i^-\rangle$ (see Eq. \ref{eq:rec1}), 
\begin{equation}
    \langle\lambda^-_{i+1}\rangle = \frac{\langle\lambda^-_{i}\rangle k+c}{1+a}.
\end{equation}
by multiplying both parts by $\lambda^-_0$, and averaging it out, we obtain a recurrent equation for the autocorrelation function
\begin{equation}
    \langle\lambda^-_{i+1}\lambda^-_0\rangle = \frac{\langle\lambda^-_{i}\lambda^-_0\rangle k+c\langle\lambda^-_0\rangle}{1+a}.
\end{equation}
by labeling $\rho^-\equiv\langle\lambda^-_0\rangle$, $g_i \equiv \langle\lambda^-_{i}\lambda^-_0\rangle$, $r = k/(1+a)$, and $b = c\rho^-/(1+an)$ we obtain a dynamic equation
\begin{equation}
    g_{i+1} = r g_i +b,
\end{equation}
which can be solved:
\begin{equation}
    g_i = r^i g_0 + b\frac{1-r^i}{1-r}.
\end{equation}
After substituting back $r$ and $b$ one obtains the equation (\ref{eq:autocorr_theory}).

\section{Mean density simulation}
\label{ap:mean_density}
While it is possible to calculate the mean density of events directly, as $N/T = 1/{\langle\tau\rangle}$ (here $N$ is the number of simulated events and $T$ is the total simulation time), the results converge very slowly due to the power-law distribution for inter-event intervals. In the paper, to speed up the convergence, we use a trick, where at each event we would first average the following inter-event intervals analytically, conditional on the post-event intensity $\lambda^+$:
\begin{equation}
\begin{split}
    \langle\tau\rangle|\lambda^+ & = \int_0^\infty p(\tau|\lambda^+)\tau d\tau\\ 
    &= \int_0^\infty P(t>\tau|\lambda^+)d\tau\\
    & = \int_0^\infty(1+a\lambda^+\tau)^{-1/a}d\tau\\
    & = \frac{1}{\lambda^+(1-a)}
\end{split}
\end{equation}
To find $\langle\tau\rangle$, we would average $\langle\tau\rangle|\lambda^+_i$ across all events in the simulation. In the main text we suggest a theoretical equation for the density (\ref{eq:density_theoretical}):
\begin{equation}
     \rho_{\text{th}} = c\frac{1-a}{1-ke^{-a}}.
\end{equation}
A comparison of the suggested $\rho_\text{th}$ to the simulated $\rho_\text{emp}$ is presented in Fig. S2. As we can see, the difference is about $30\%$ at the higher end, and $20\%$ at the lower end, while being less than $5\%$ for either $k<0.5$ or $a<0.5$. Given that the equation was constructed with the sole purpose of satisfying edge cases, we consider this to be a remarkably good result.
\begin{figure}
    \centering
    \includegraphics[width=0.5\linewidth]{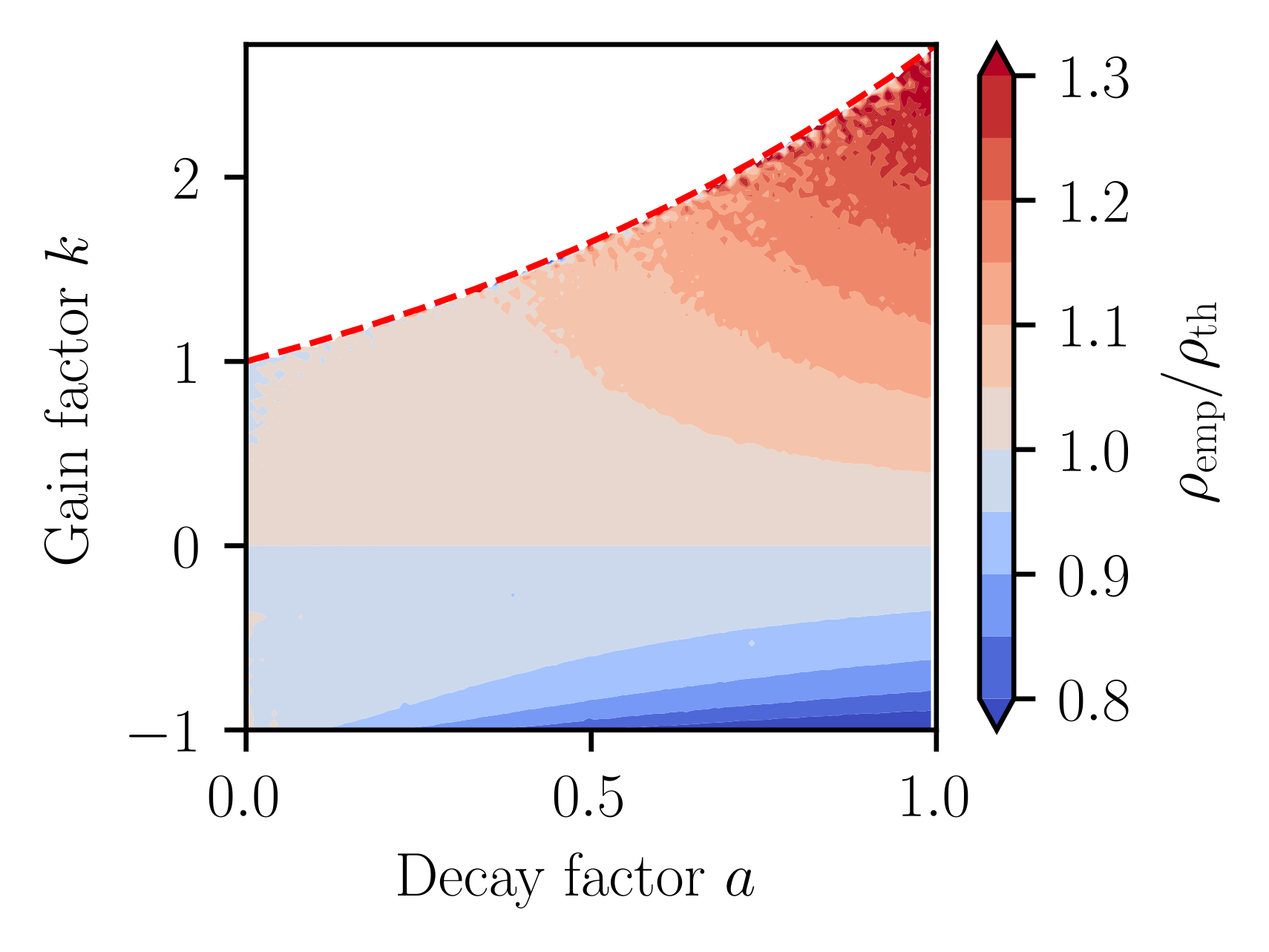}
    \caption{\label{fig:s2} The ratio of the simulated event density $\rho_{\text{emp}}$ to theoretical $\rho_{\text{th}}$, expressed by Eq. (\ref{eq:density_theoretical}). For the simulation, we assumed $c = 1$, and $N_{\text{events}}=10^5$.}
\end{figure}

\section{Tail of the conditional probability $p(\text{event at }  t_2|\text{event at } t_1)$}
\label{ap:condition_prob_tail}

\begin{figure}
\centering
\includegraphics[width=0.67\linewidth]{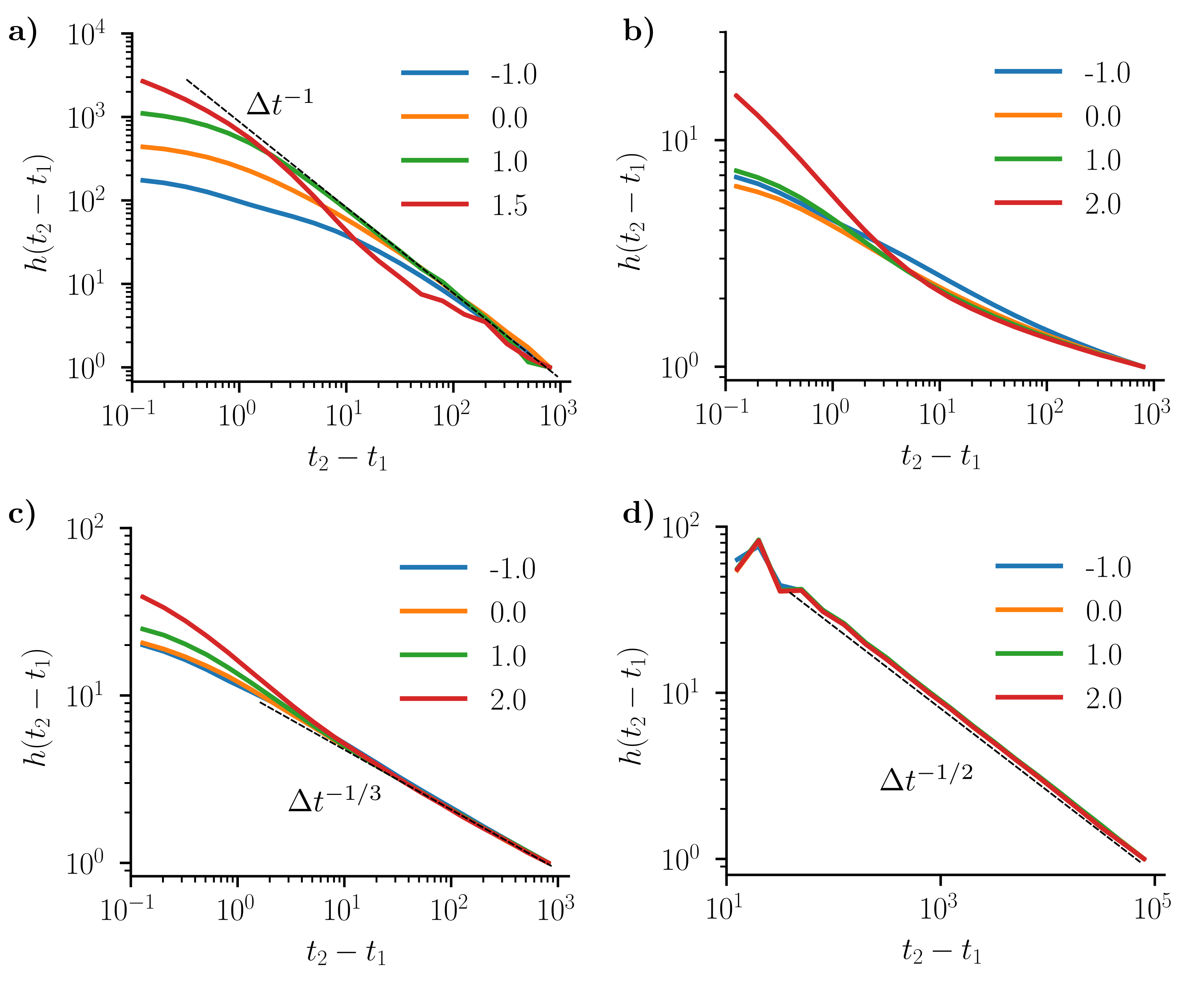}
\caption{\label{fig:autocorr_tail}The simulation of the rescaled conditional probability $h(t_2-t_1)=p(\text{event at } t_2|\text{event at } t_1)-\rho$, $h(t_2-t_1)/h_{10^3}$ vs $\Delta t = t_2 - t_1$. The number of simulated events was taken to be $N_{\text{events}} = 10^7$. a) $a = 0.5$. b) $a = 1$. c) $a = 1.5$. d) $a = 2$.}

\label{fig:s2}
\end{figure}

Here we provide a detailed illustration that the {\bf{tail}} of the conditional probability density $h(t_2-t_1)=p(\text{event at } t_2|\text{event at } t_1)-\rho$ as a function of $\Delta t = t_2-t_1$, has the same asymptotic behavior for different values of $k$. Fig. \ref{fig:autocorr_tail} provides the rescaled $h(t_2-t_1)/h(\Delta t=1000)$ for four different values of $a$: $a = \{0.5, 1,  1.5, 2\}$. For a renewal process with a power law distributed IEI $\tau\sim \tau^{-\alpha_{\text{IEI}}}$, it has been derived that auto-covariance behaves as $h(\Delta t)\sim\Delta t^{-\alpha_{\text{cov}}}$, where $\alpha_{\text{cov}} = 2 - \alpha_{\text{IEI}}$ for $\alpha_{\text{IEI}} < 1$ and $\alpha_{\text{cov}} = \alpha_{\text{IEI}}-2$ for $\alpha_{\text{IEI}} > 1$. As we can see, the tails of the distributions match the predicted slope of $\alpha_{\text{cov}} = 1 - 1/a$ renewal processes ($k = 0$) as well as for the general model ($k \neq 0$).

\section{Validation and Residuals}
\label{ap:validation}

Here we discuss model validation.
Since our model predicts the probability distribution of IEIs rather than their exact values, standard cross-validation techniques are not directly applicable. Instead, one way to assess model validity is by comparing simulated and empirical distributions using the Kolmogorov-Smirnov (KS) test. For example, one can compare the distribution of IEIs, or the distribution of event counts within a chosen time window. While these methods can falsify the model if discrepancies arise, they do not reveal where or why the model fails. For example, one could gain specific insights into how to modify the model if one knew that it tends not to reproduce the data well after long periods of inactivity or after closely spaced events.

An alternative approach is based on the Probability Integral Transform (PIT) residuals \cite{dawid_present_1984}. The key idea here is that for any probability distribution with cumulative function $P(x)$, the transformed variable $y = P(x)$ should follow a uniform distribution on $\left[0,1\right]$. Applying this to our case, the cumulative probability of each observed IEI $\tau_i$, denoted $P_i(\tau_i)$, should be uniformly distributed if the model indeed describes the data.

Within the PIT framework, a basic model validation check would be to see whether the empirical distribution of $P_i(\tau_i)$ matches a uniform distribution using a KS test. However, this approach still has the limitation that if the model exhibits systematic biases in different regimes (e.g., underestimating probability of short intervals after large gaps while overestimating the same probability after short gaps), these effects may cancel out in aggregate.

To address concerns like the above, we can refine the test by conditioning it on the previous event interval. Specifically for this example, we can examine the conditional distribution $P_i(\tau_i|\tau_{i-1})$, which should also be uniform. This can be assessed by plotting histograms of the empirical joint distribution $P_i(\tau_i,\tau_{i-1})$ and looking for deviations from uniformity. This method provides a more granular assessment of model fit, revealing specific regimes where the model's assumptions may break down. For example, we consider a model with a reset function 
\begin{equation}
    f(\lambda) = k\frac{\lambda}{1+\lambda}.
    \label{eq:slow_start_rec}
\end{equation}
An important feature of the model is $f(\lambda) \rightarrow 0$ as $\lambda\rightarrow 0$. We call models with such property {\it slow start} models. In such models, if an event happen after a long period of inactivity (i.e. $\lambda$ is small), the next IEI is likely to be also large, since the new post-event intensity $\lambda$ is still small. This is unlike the linear reset model, where the intensity resets back to around $\lambda = c$ after long inactivity periods. In other respects the slow start model can behave similarly to the linear reset model, as both can feature power law IEIs and positive correlations between inter-event intervals. 

Due to long periods of inactivity likely to be followed by a large IEI, one expects that if the data has been generated by the slow start model, but we fit our linear model, the model would underestimate the number of long intervals after long intervals. 

Empirical simulation indeed confirms this intuition. The results for the $k = 3.2$, and $a = 1$, and $N_{\text{events}}=2000$. In Fig. \ref{fig:s_validation}, we can see that while the cumulative PIT residuals are consistent with the uniform distribution (Fig. \ref{fig:s_validation}b), the conditional probability distribution $P_i(\tau_i|\tau_{i-1})$ is highly skewed towards higher values (Fig. \ref{fig:s_validation}c), which means that the data indeed contains much more longer intervals than expected.

\begin{figure}
    \centering
\includegraphics[width=0.8\linewidth]{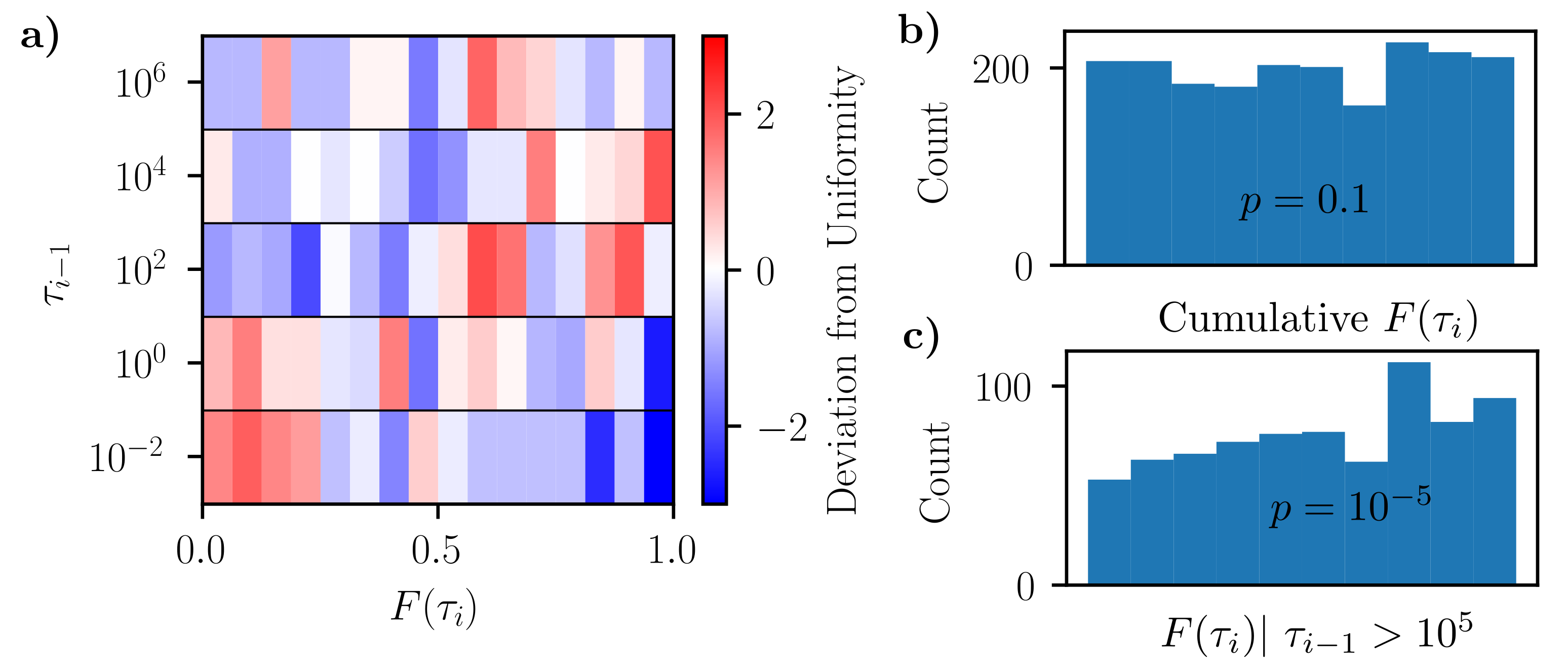}
    \caption{\label{fig:s_validation} Model validation using conditional PIT residuals to detect model mismatch.  Data were generated from a "slow start" model ($f(\lambda) = k \lambda/(1+\lambda)$, Eq. \eqref{eq:slow_start_rec}) with $N_{\text{event}} = 2000$, $k = 3.2$, and $a = 1$. a) Deviations from uniformity for the conditional PIT residuals, $F(\tau_i|\tau_{i-1})$. b) The marginal distribution of all PIT residuals is consistent with a uniform distribution (Kolmogorov-Smirnov test, $p=0.1$) c) The conditional distribution of residuals following long intervals ($\tau_{i-1} > 10^5$) is strongly skewed towards 1 (KS test, $p=10^{-5}$), showing that the misspecified model fails by underestimating the probability of long IEIs after a long period of inactivity.}
\end{figure}

This idea extends naturally to any conditioning variable of interest that can be calculated using the event history.

\section{Analysis of the Negative Decay Parameter Regime ($a<0$)}
\label{ap:negative_a}
\begin{figure}
    \centering
    \includegraphics[width=0.8\linewidth]{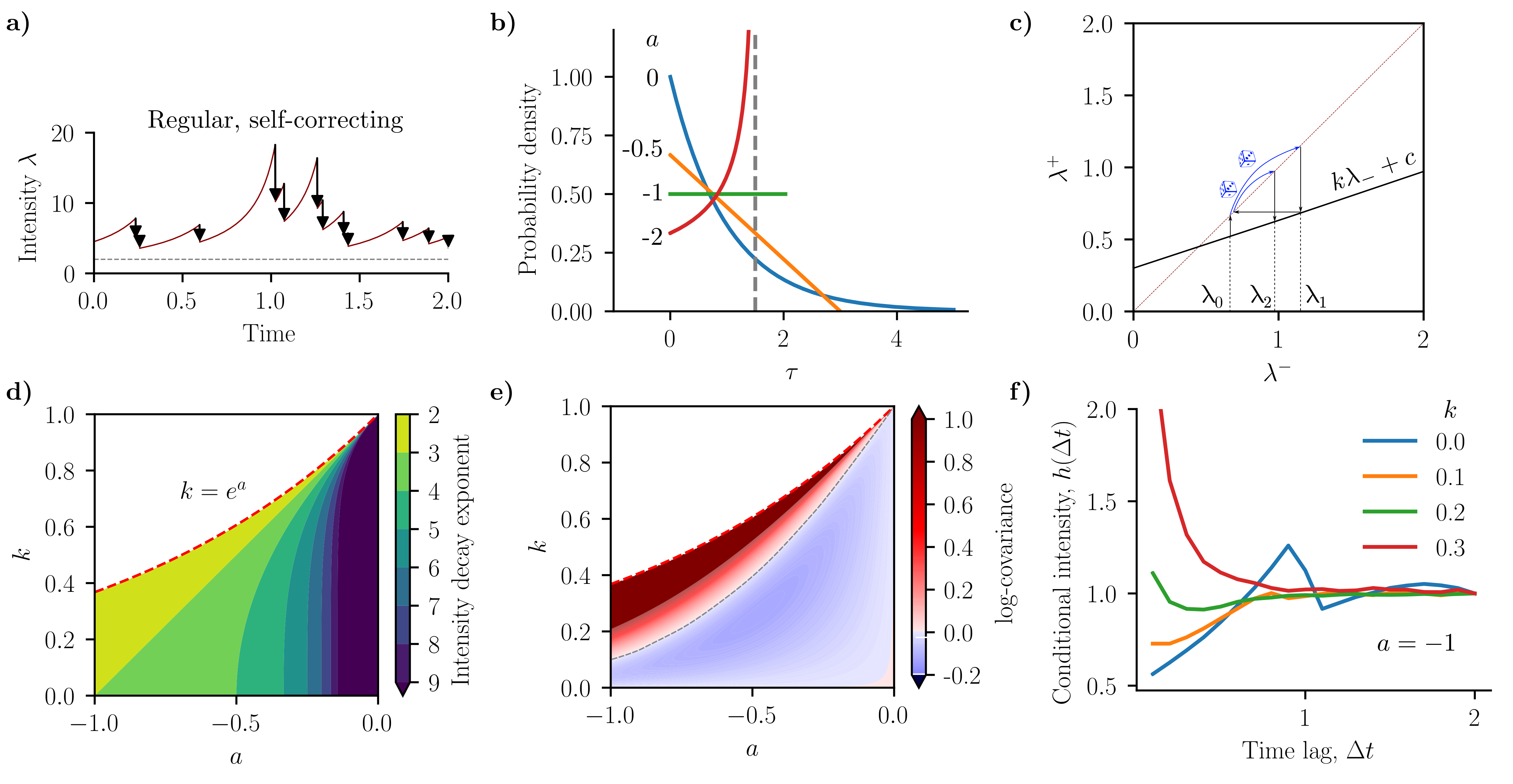}
    \caption{
    \label{fig:negative_a} 
    Characteristics of the model in the negative decay regime ($a<0$), which produces regular dynamics instead of bursts. 
    \textbf{a)} An example intensity trajectory for $a = -0.5$ and $k = 0.5$, showing regular, self-correcting spiking. 
    \textbf{b)} Cobweb plot illustrating the event-to-event intensity map, $\lambda^+ = k\lambda^-+c$. 
    \textbf{c)} Inter-event interval (IEI) probability distributions for the constant reset model ($k=0$) at different values of $a$, showing characteristic hard cutoffs. The reset value $\lambda_0=1/(1-a)$ was chosen to preserve the mean IEI across parameters. 
    \textbf{d)} Power-law exponent of the stationary intensity distribution as a function of $a$ and $k$. The red dashed line indicates the stability boundary $k=e^a$.
    \textbf{e)} Log-covariance of neighboring IEIs, showing parameter regimes of both positive (red) and negative (blue) correlation. The grey dashed line indicates zero covariance. 
    \textbf{f)} The conditional intensity $h(\Delta t)$ for $a=-1$, showing oscillatory behavior for small $k$, which is characteristic of regular spiking.
}
\end{figure}

While the main text focused on the positive decay parameter ($a > 0$) to model bursty dynamics, our framework remains well-defined for $a < 0$. In this regime, the intensity grows between events according to $\dot{\lambda} = |a|\lambda^2$ (Fig. \ref{fig:negative_a}a), leading to qualitatively different, regular rather than bursty event patterns. This appendix analyzes the key properties of this negative decay regime.

\paragraph{Constant reset model.}
We begin by analyzing the constant reset model, where the intensity is reset to a fixed value $\lambda_0$ after each event. Although the analytical form of the IEI probability distribution (Eqs. \eqref{eq:CDF1} and \eqref{eq:inter_event1}) remains valid, the behavior is qualitatively different for $a < 0$. The PDF of the IEI distribution now has a hard cutoff and becomes zero for $\tau > \tau_{\text{max}} = (|a|\lambda_0)^{-1}$. The shape of the distribution depends qualitatively on the magnitude of the decay parameter  (Fig. \ref{fig:negative_a}b):
\begin{enumerate}
    \item For $|a| < 1$ the PDF of IEIs is a decreasing function of $\tau$ and becomes zero at $\tau = \tau_{\text{max}}$.
    \item For $|a| > 1$ the PDF of IEIs is an increasing function of $\tau$ and goes to infinity at $\tau \rightarrow \tau_{\text{max}}$.
    \item For $|a| = 1$ the PDF is a uniform distribution.
\end{enumerate}

In all cases, the $a < 0$ behavior results in event sequences that are more regular than a Poisson process, as short inter-event intervals are suppressed compared to intervals corresponding to the mean density. 

\paragraph{Linear reset model.}
Next, we consider the linear reset model defined by Eq. \eqref{eq:iterative}. We analyze this model in the event-domain representation. Since the derivation of the Kesten form for the post-event (Eq. \eqref{eq:kesten_recursion}) and pre-event (Eq. \ref{eq:kesten_recursion__}) intensities did not depend on the sign of $a$, these equations still hold. The picture of the dynamics in the event domain remain similar, with one critical difference: the stochastic decay between events is replaced by a stochastic increment (see Fig. \ref{fig:negative_a}c). Consequently, for the process to be stable, the deterministic reset jump $\lambda^- \rightarrow\lambda^+$ must be negative for large $\lambda$ to counteract the stochastic growth, which requires $k < 1$. Furthermore, for the intensity to remain non-negative, we require $k \ge 0$, as a negative k could produce an unphysical negative intensity given the unbounded stochastic increment.

The stability condition, $|k|<e^a$ remains unchanged, as its derivation is independent of the sign of $a$ (see. Appendix\ref{ap:stability_analysis}). One can infer from the cobweb diagram of the dynamics that the post-event intensity is bounded from below by the point $\lambda^+_{\text{min}} = c/(1-k)$. This minimum intensity, in turn, imposes a hard cutoff on the IEI distribution at $\tau_{\text{max}} = (|a|\lambda^+_{\text{min}})^{-1}$.

As it turns out, the stationary intensity distribution is a power law for all valid $k$, with an exponent determined by Eq. \eqref{eq:transend} (Fig. \ref{fig:negative_a}d). At first glance, this result may seem puzzling: it is not immediately obvious how a process with regular event timings can exhibit a power-law intensity distribution. The resolution to this apparent paradox lies in recognizing that only the event times are observable, not the intensity itself. The explosive growth of intensity between events allows for a heavy-tailed distribution of intensity values, even if the resulting event sequence is regular. This raises an important methodological question: how can one infer the tail of the intensity distribution from event data alone? A rigorous method for this is an open question that we leave for future research.

The correlation structure in this regime is non-trivial, arising from a competition between interval-duration feedback and intensity marginalization. The origin of these competing effects can be understood by analyzing the joint distribution of successive inter-event intervals, $p(\tau_i, \tau_{i+1})$. First, if we stratify this distribution and consider the conditional probability $p(\tau_i, \tau_{i+1}|\lambda^+_i)$, a negative correlation emerges. For a fixed starting intensity $\lambda^+_i$, a stochastically longer $\tau_i$ allows the intensity to grow to a larger value $\lambda^-_{i+1}$. 
The reset rule maps this to a higher subsequent intensity $\lambda^+_{i+1}$, which in turn generates a statistically shorter following interval $\tau_{i+1}$. This mechanism, which we term \textbf{interval-duration feedback}, creates a direct negative correlation between adjacent IEIs.

To recover the full joint distribution, $p(\tau_i, \tau_{i+1})$, we must marginalize over the distribution of the latent intensity variable, $p(\lambda^+_i)$. A higher value of $\lambda^+_i$ systematically shifts the conditional distribution $p(\tau_i, \tau_{i+1}|\lambda^+_i)$ towards smaller values in both $\tau_i$ and $\tau_{i+1}$. Averaging over all possible starting intensities therefore induces a positive correlation. We call this the \textbf{intensity marginalization effect}.

Simulations show  (Fig. \ref{fig:negative_a}e) that for small k, the interval-duration feedback dominates, and the net correlation is negative. As $k$ increases towards the stability boundary $k_{\text{cr}}=e^a$, the distribution of the starting intensity $p{(\lambda^+_i)}$ becomes significantly heavy tailed. The increased variance from this marginalization effect then dominates the conditional negative feedback, resulting in a net positive correlation.

In the time domain, this competition manifests as an oscillatory autocorrelation function for small $k$, characteristic of regular spiking. As $k$ increases, this is replaced by a gradual increase in positive correlations at short time lags (Fig. \ref{fig:negative_a}f).

In conclusion, the linear reset process with $a<0$ exhibits a dual nature: it is regular on a large time scale, with a strict upper bound on IEI durations, while exhibiting strong, transient spikes in intensity on a small time scale. The competition between large-scale regularity and small-scale intensity spikes creates a complex correlation structure. This suggests the need for analytical tools that can characterize the behavior of such processes at different scales, which could be a topic for future research.

\section{Power law inter-event intervals with Nonlinear Hawkes processes}
\label{ap:pl_hawkes}

This section demonstrates the construction of a nonlinear Hawkes process capable of generating inter-event interval (IEI) distributions that follow a power law. The central insight is that to produce such a distribution, the process's intensity, $\lambda(t)$, must decay as $1/t$ in the time following an event.

A nonlinear Hawkes process models the intensity, $\lambda(t)$, as a nonlinear function, $f$, of a hidden state variable $\nu(t)$, which we call the "tension" :
\begin{equation}
    \lambda = g(\nu)
\end{equation}
The tension $\nu(t)$ integrates the history of past events $\{t_i|t_i<t\}$ through a memory kernel $\phi(t)$:
\begin{equation}
    \nu = \nu_0+\sum_{t_i}\phi(t-t_i),
\end{equation}
where $\nu_0$ is a baseline tension.

For mathematical tractability, we choose an exponential memory kernel $\phi(t)= Ae^{-t}$. This choice is convenient because it endows the tension $\nu(t)$ with simple dynamics: between events, the tension decays exponentially, following the first-order differential equation $\dot{\nu}=-\nu$. After each event, the tension jumps by $A$.  

Our goal is to find the form of the function $f$ that transforms the exponential decay of tension $\nu(t)$ into the desired $1/t$ decay of intensity $\lambda(t)$. 

To achieve this, let us assume $\nu = e^{-t}$. The $1/t$ intensity decay scaling then suggests $f(e^{-t}) = (\alpha-1)/t$, which yields 
\begin{equation}
    g(\nu) = -\frac{\alpha-1}{\ln(\nu)}
\end{equation}
This function has a singularity at $\nu = 1$, where the intensity becomes infinite. To remedy this - while preserving the behavior at small $\nu$ - we modify the function so that the argument of the logarithm never reaches 1. A well behaved choice is
\begin{equation}
    g(\nu) = -\frac{\alpha-1}{\ln(\nu/(\nu+1))},
    \label{eq:nlHawkes_fun}
\end{equation}

This function behaves as $g(\nu)\propto (\alpha-1)  \nu$ for large $\nu$, while still behaving as $-(\alpha-1)/\ln(\nu)$ for small $\nu$. 

We can confirm that our choice of $g(\nu)$ produces the correct intensity decay. Substituting $\nu = C e^{-t}$ into regularized $g(\nu)$ yields:
\begin{equation}
    \lambda = f(\nu(t)) = \frac{\alpha-1}{t+\ln {(C^{-1}+e^{-t})}}\propto \frac{\alpha-1}{t},
\end{equation}
which confirms our choice of the function $f$. Such choice of $f$ is expected to yield IEI distribution with a power-law tail $p(\tau)\propto \tau^{-\alpha}$, which is confirmed empirically in Fig. S1a. 

The empirical distribution for IEIs confirms the correct power-law behavior of the IEI exponent (Fig. S1a).

\begin{figure}
\centering
\includegraphics[width=0.8\linewidth]{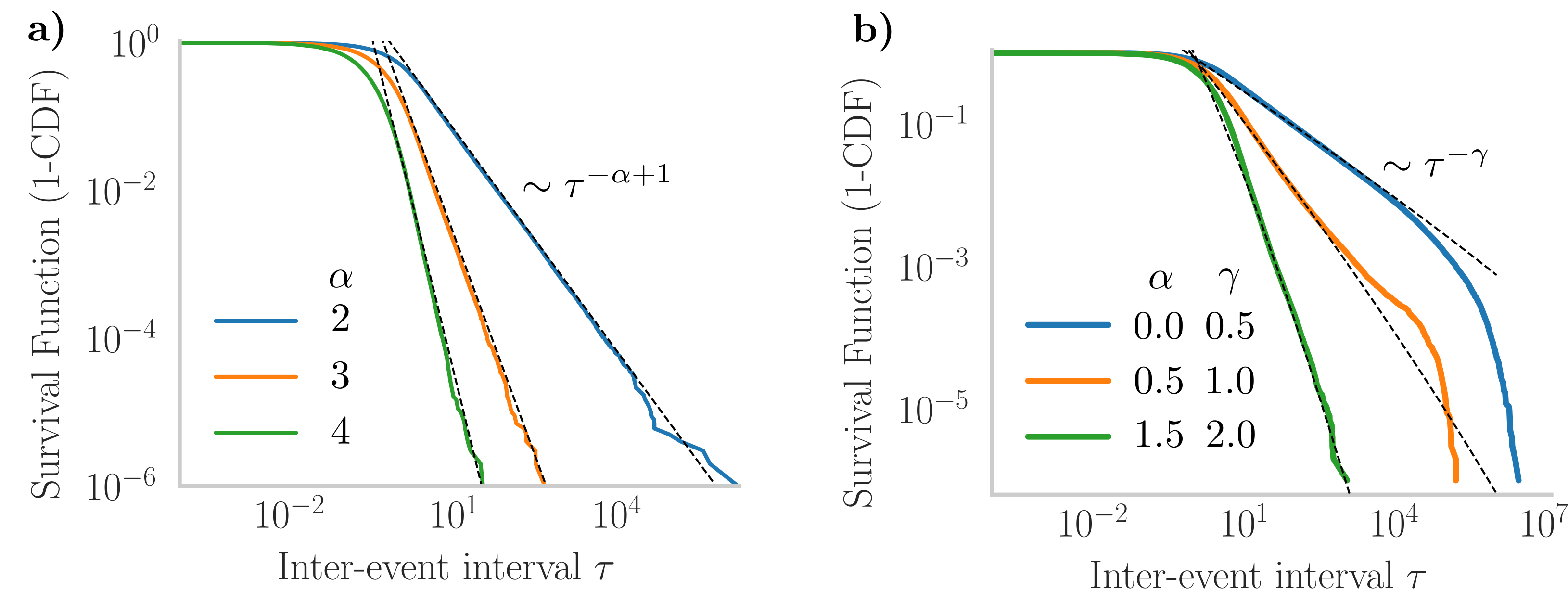}
\caption{Empirical survival function for nonlinear Hawkes process and Cox process. a) Survival function for nonlinear Hawkes process with the kernel $\phi(t) = 0.2e^{-t}$, and tension-intensity function (\ref{eq:nlHawkes_fun}) for different values $\alpha$. b) Survival function for the Cox process, with $\beta = 1.5$}
\label{fig:hawkes_cox_power_law}
\end{figure}

\section{Power law inter-event intervals with a Cox process}
\label{ap:pl_cox}

Here we provide a possible way to generate a power law IEI distribution with a Cox process. As far as we know, a pure power-law distribution in Cox processes have not been obtained In Cox processes intensity fluctuates independently of event occurrences. The intuition for the construction of the process is the following. We would like our process to generate many long inter-event intervals, which means that intensity must fluctuate to very low values. Also, intensity must spend enough time at these low values, so events actually occur.

Here is possible recipe that includes both requirements. The process is constructed by dividing the timeline into epochs, where each epoch has a constant Poisson intensity.

The model is defined by the following rules for the stochastic intensity $\lambda(t)$:
\begin{enumerate}
    \item At the beginning of an epoch, an intensity $\lambda$ is drawn from a prior probability distribution $p(\lambda)$ given by a power law with an upper cutoff $\lambda_{\text{top}}$:
    \begin{equation}
        p(\lambda) = C_p \lambda^\alpha, \quad \text{for } \lambda \in (0, \lambda_{\text{top}}].
        \label{eq:prior_lambda}
    \end{equation}
    Here, $\alpha > -1$ is a parameter and $C_p = (\alpha+1)/\lambda_{\text{top}}^{\alpha+1}$ is the normalization constant.

    \item The drawn intensity $\lambda$ is held constant for a duration $T_\lambda$, which is inversely related to the intensity itself:
    \begin{equation}
        T_\lambda = c \lambda^{-\beta},
        \label{eq:duration_lambda}
    \end{equation}
    where $c$ is a constant of proportionality and $\beta > 1$ is a key parameter ensuring that low-intensity epochs are long-lived.

    \item After the duration $T_\lambda$ has passed, a new intensity is drawn independently from the same distribution $p(\lambda)$, and the process repeats.
\end{enumerate}
The resulting point process is a doubly stochastic Poisson process where the waiting time $\tau$ between consecutive events, for a fixed intensity $\lambda$, follows an exponential distribution with survival function $S(\tau|\lambda) = e^{-\tau\lambda}$.

To find the overall IEI distribution, we must average over $\lambda$. Crucially, we must use the distribution of $\lambda$ as experienced at the time of an event, $p_{\text{event}}(\lambda)$, not the prior distribution $p(\lambda)$, due to the inspection paradox: an event is more likely to occur during a high-intensity epoch. The average number of events, $N_\lambda$, in an epoch with intensity $\lambda$ is $N_\lambda = \lambda T_\lambda = c \lambda^{1-\beta}$. The probability density $p_{\text{event}}(\lambda)$ is thus proportional to the prior $p(\lambda)$ weighted by the number of events contributed, $N_\lambda$:
\begin{equation}
    p_{\text{event}}(\lambda) \propto p(\lambda) \times N_\lambda \propto (\lambda^\alpha)(\lambda^{1-\beta}) = \lambda^{\alpha-\beta+1}.
\end{equation}
We define the exponent $\gamma \equiv \alpha - \beta + 1$. For this distribution to be normalizable on $(0, \lambda_{\text{top}}]$, we require $\gamma > -1$. The normalized distribution is:
\begin{equation}
    p_{\text{event}}(\lambda) = C_\gamma \lambda^\gamma = \frac{\gamma+1}{\lambda_{\text{top}}^{\gamma+1}} \lambda^\gamma, \quad \text{for } \lambda \in (0, \lambda_{\text{top}}].
    \label{eq:pevent_lambda}
\end{equation}
With this event-conditioned intensity distribution, the overall survival function for the IEI, $S(\tau) = P(t > \tau)$, is obtained by integrating the conditional survival function over all possible values of $\lambda$:
\begin{equation}
    S(\tau) = \int_0^{\lambda_{\text{top}}} S(\tau|\lambda) \ p_{\text{event}}(\lambda) \ d\lambda = \int_0^{\lambda_{\text{top}}} e^{-\tau\lambda} \left( \frac{\gamma+1}{\lambda_{\text{top}}^{\gamma+1}} \lambda^\gamma \right) d\lambda.
    \label{eq:survival_integral}
\end{equation}
This integral is valid under the assumption that the intensity is constant over a typical IEI, which holds for the low-$\lambda$ epochs that generate the power-law tail since the mean IEI ($1/\lambda$) is much shorter than the epoch duration ($c\lambda^{-\beta}$) for small $\lambda$ when $\beta > 1$.

To analyze the asymptotic behavior for large $\tau$, we perform a change of variables, letting $x = \tau\lambda$, so $d\lambda = dx/\tau$. The integral in Eq.~\eqref{eq:survival_integral} becomes:
\begin{align}
    S(\tau) &= \frac{\gamma+1}{\lambda_{\text{top}}^{\gamma+1}} \int_0^{\tau\lambda_{\text{top}}} e^{-x} \left(\frac{x}{\tau}\right)^\gamma \frac{dx}{\tau} 
    = \frac{\gamma+1}{(\tau\lambda_{\text{top}})^{\gamma+1}} \int_0^{\tau\lambda_{\text{top}}} e^{-x} x^\gamma dx.
\end{align}
The integral is the lower incomplete gamma function. For large $\tau$ (specifically, $\tau \gg 1/\lambda_{\text{top}}$), its upper limit $\tau\lambda_{\text{top}} \to \infty$, and the integral approaches the complete Gamma function, $\Gamma(\gamma+1)$. In this limit, the survival function has the asymptotic form:
\begin{equation}
    S(\tau) \approx \frac{(\gamma+1)\Gamma(\gamma+1)}{(\lambda_{\text{top}})^{\gamma+1}} \tau^{-(\gamma+1)},
\end{equation}
which shows that the tail of the survival function follows a power law, $S(\tau) \propto \tau^{-(\gamma+1)}$. The PDF of the IEI, $P(\tau) = -dS(\tau)/d\tau$, is found by differentiating this expression, yielding $P(\tau) \propto \tau^{-(\gamma+2)}$.

Finally, by substituting the definition $\gamma = \alpha - \beta + 1$ into the exponent, we obtain the final expression for the power-law tail of the IEI distribution:
\begin{equation}
    P(\tau) \propto \tau^{-\mu} \quad \text{with exponent} \quad \mu = \alpha - \beta + 3.
    \label{eq:final_exponent}
\end{equation}
This result holds for $\tau \gg 1/\lambda_{\text{top}}$ and requires the condition $\alpha - \beta + 1 > -1$ for the distributions to be normalizable. Empirical simulation with $\beta = 1.5$ are consistent with the predicted slope, although deviations can be observed in the tails of the distribution due to the correlations between events.

\section{Mapping a Non-Linear Hawkes Process with an Exponential Kernel to a General Markovian Point Process}
\label{ap:Hawkes_markov_mapping}

Here, we demonstrate that a non-linear Hawkes process with an exponential memory kernel can be mapped onto a specific instance of the general Markovian framework presented in this paper.
Consider a non-linear Hawkes process where the memory kernel is a single exponential, $\phi(t) = A e^{-\alpha t}$. The dynamics of the latent tension variable, $\nu$, between events can be described by the ordinary differential equation:
\begin{equation}
    \dot{\nu} = -\alpha \nu.
\end{equation}
Let the instantaneous event rate, $\lambda(t)$, be an invertible, non-linear function of the tension, $\lambda = g(\nu)$. Taking the time derivative of $\lambda$ and applying the chain rule yields $\dot{\lambda} = \dot{\nu}g'(\nu)$. By substituting the expression for $\dot{\nu}$ and then expressing $\nu$ in terms of $\lambda$ (i.e. $\nu = f^{-1} (\lambda)$), we obtain the dynamics equation for the event rate:
\begin{equation}
    \dot{\lambda} = -\alpha g^{-1}(\lambda) g'\left(g^{-1}(\lambda)\right).
\end{equation}
This equation defines the decay function $\dot{\lambda} = F(\lambda)$ for this equivalent Markov process. Next, we derive the corresponding reset function, $f(\lambda^-)$. In a Hawkes process, the tension instantaneously jumps by a constant amount upon occurance of each event, $\nu^+ = \nu^- + A$.  We can map this rule into the intensity domain:
\begin{equation}
\begin{split}
    \lambda^+ &= g(\nu^+) = g(\nu^- +A)\\
    & =g\left(g^{-1}(\lambda^-)+A\right)
\end{split}
\end{equation}
This gives the reset function $f(\lambda^-) = g\left(g^{-1}(\lambda^-)+A\right)$. This derivation shows that any non-linear Hawkes process with an exponential kernel and an invertible tension-intensity function can be expressed as a Markovian point process with a specific non-linear decay law and reset function. This demonstrates that such Hawkes processes are a subclass of the general framework we consider.

\printbibliography

@article{kanazawa_asymptotic_2023,
	title = {Asymptotic solutions to nonlinear {Hawkes} processes: {A} systematic classification of the steady-state solutions},
	volume = {5},
	copyright = {https://creativecommons.org/licenses/by/4.0/},
	issn = {2643-1564},
	shorttitle = {Asymptotic solutions to nonlinear {Hawkes} processes},
	url = {https://link.aps.org/doi/10.1103/PhysRevResearch.5.013067},
	doi = {10.1103/physrevresearch.5.013067},
	language = {en},
	number = {1},
	urldate = {2024-09-17},
	journal = {Physical Review Research},
	author = {Kanazawa, Kiyoshi and Sornette, Didier},
	month = jan,
	year = {2023},
	note = {Publisher: American Physical Society (APS)},
}

@article{karsai_universal_2012,
	title = {Universal features of correlated bursty behaviour},
	volume = {2},
	copyright = {https://creativecommons.org/licenses/by-nc-sa/3.0/},
	issn = {2045-2322},
	url = {https://www.nature.com/articles/srep00397},
	doi = {10.1038/srep00397},
	language = {en},
	number = {1},
	urldate = {2024-09-17},
	journal = {Scientific Reports},
	author = {Karsai, Márton and Kaski, Kimmo and Barabási, Albert-László and Kertész, János},
	month = may,
	year = {2012},
	note = {Publisher: Springer Science and Business Media LLC},
}

@article{kanazawa_ubiquitous_2021,
	title = {Ubiquitous {Power} {Law} {Scaling} in {Nonlinear} {Self}-{Excited} {Hawkes} {Processes}},
	volume = {127},
	url = {https://link.aps.org/doi/10.1103/PhysRevLett.127.188301},
	doi = {10.1103/PhysRevLett.127.188301},
	number = {18},
	urldate = {2024-09-22},
	journal = {Physical Review Letters},
	author = {Kanazawa, Kiyoshi and Sornette, Didier},
	month = oct,
	year = {2021},
	note = {Publisher: American Physical Society},
	pages = {188301},
}

@article{barabasi_origin_2005,
	title = {The origin of bursts and heavy tails in human dynamics},
	volume = {435},
	copyright = {2005 Macmillan Magazines Ltd.},
	issn = {1476-4687},
	url = {https://www.nature.com/articles/nature03459},
	doi = {10.1038/nature03459},
	language = {en},
	number = {7039},
	urldate = {2024-09-22},
	journal = {Nature},
	author = {Barabási, Albert-László},
	month = may,
	year = {2005},
	note = {Publisher: Nature Publishing Group},
	pages = {207--211},
}

@inproceedings{henderson_modelling_2001,
	address = {Ottawa Canada},
	title = {Modelling user behaviour in networked games},
	isbn = {978-1-58113-394-3},
	url = {https://dl.acm.org/doi/10.1145/500141.500175},
	doi = {10.1145/500141.500175},
	language = {en},
	urldate = {2024-09-22},
	booktitle = {Proceedings of the ninth {ACM} international conference on {Multimedia}},
	publisher = {ACM},
	author = {Henderson, Tristan and Bhatti, Saleem},
	month = oct,
	year = {2001},
	pages = {212--220},
}

@article{zhou_role_2008,
	title = {Role of activity in human dynamics},
	volume = {82},
	issn = {0295-5075, 1286-4854},
	url = {https://iopscience.iop.org/article/10.1209/0295-5075/82/28002},
	doi = {10.1209/0295-5075/82/28002},
	language = {en},
	number = {2},
	urldate = {2024-09-28},
	journal = {EPL (Europhysics Letters)},
	author = {Zhou, T. and Kiet, H. A. T. and Kim, B. J. and Wang, B.-H. and Holme, P.},
	month = apr,
	year = {2008},
	pages = {28002},
}

@article{kleinberg_bursty_nodate,
	title = {Bursty and {Hierarchical} {Structure} in {Streams}},
	language = {en},
	author = {Kleinberg, Jon},
}

@article{gabaix_theory_2003,
	title = {A theory of power-law distributions in financial market fluctuations},
	volume = {423},
	copyright = {http://www.springer.com/tdm},
	issn = {0028-0836, 1476-4687},
	url = {https://www.nature.com/articles/nature01624},
	doi = {10.1038/nature01624},
	language = {en},
	number = {6937},
	urldate = {2024-10-01},
	journal = {Nature},
	author = {Gabaix, Xavier and Gopikrishnan, Parameswaran and Plerou, Vasiliki and Stanley, H. Eugene},
	month = may,
	year = {2003},
	pages = {267--270},
}

@article{corral_long-term_2004,
	title = {Long-{Term} {Clustering}, {Scaling}, and {Universality} in the {Temporal} {Occurrence} of {Earthquakes}},
	volume = {92},
	copyright = {http://link.aps.org/licenses/aps-default-license},
	issn = {0031-9007, 1079-7114},
	url = {https://link.aps.org/doi/10.1103/PhysRevLett.92.108501},
	doi = {10.1103/PhysRevLett.92.108501},
	language = {en},
	number = {10},
	urldate = {2024-10-01},
	journal = {Physical Review Letters},
	author = {Corral, Álvaro},
	month = mar,
	year = {2004},
	pages = {108501},
}

@article{malamud_forest_1998,
	title = {Forest {Fires}: {An} {Example} of {Self}-{Organized} {Critical} {Behavior}},
	volume = {281},
	issn = {0036-8075, 1095-9203},
	shorttitle = {Forest {Fires}},
	url = {https://www.science.org/doi/10.1126/science.281.5384.1840},
	doi = {10.1126/science.281.5384.1840},
	language = {en},
	number = {5384},
	urldate = {2024-10-01},
	journal = {Science},
	author = {Malamud, Bruce D. and Morein, Gleb and Turcotte, Donald L.},
	month = sep,
	year = {1998},
	pages = {1840--1842},
}

@article{touati_origin_2009,
	title = {Origin and {Nonuniversality} of the {Earthquake} {Interevent} {Time} {Distribution}},
	volume = {102},
	copyright = {http://link.aps.org/licenses/aps-default-license},
	issn = {0031-9007, 1079-7114},
	url = {https://link.aps.org/doi/10.1103/PhysRevLett.102.168501},
	doi = {10.1103/PhysRevLett.102.168501},
	language = {en},
	number = {16},
	urldate = {2024-10-01},
	journal = {Physical Review Letters},
	author = {Touati, Sarah and Naylor, Mark and Main, Ian G.},
	month = apr,
	year = {2009},
	pages = {168501},
}

@article{beggs_neuronal_2003,
	title = {Neuronal {Avalanches} in {Neocortical} {Circuits}},
	volume = {23},
	copyright = {https://creativecommons.org/licenses/by-nc-sa/4.0/},
	issn = {0270-6474, 1529-2401},
	url = {https://www.jneurosci.org/lookup/doi/10.1523/JNEUROSCI.23-35-11167.2003},
	doi = {10.1523/JNEUROSCI.23-35-11167.2003},
	language = {en},
	number = {35},
	urldate = {2024-10-01},
	journal = {The Journal of Neuroscience},
	author = {Beggs, John M. and Plenz, Dietmar},
	month = dec,
	year = {2003},
	pages = {11167--11177},
}

@article{sorribes_origin_2011,
	title = {The {Origin} of {Behavioral} {Bursts} in {Decision}-{Making} {Circuitry}},
	volume = {7},
	issn = {1553-7358},
	url = {https://dx.plos.org/10.1371/journal.pcbi.1002075},
	doi = {10.1371/journal.pcbi.1002075},
	language = {en},
	number = {6},
	urldate = {2024-10-01},
	journal = {PLoS Computational Biology},
	author = {Sorribes, Amanda and Armendariz, Beatriz G. and Lopez-Pigozzi, Diego and Murga, Cristina and De Polavieja, Gonzalo G.},
	editor = {Sporns, Olaf},
	month = jun,
	year = {2011},
	pages = {e1002075},
}

@article{proekt_scale_2012,
	title = {Scale invariance in the dynamics of spontaneous behavior},
	volume = {109},
	issn = {0027-8424, 1091-6490},
	url = {https://pnas.org/doi/full/10.1073/pnas.1206894109},
	doi = {10.1073/pnas.1206894109},
	language = {en},
	number = {26},
	urldate = {2024-10-01},
	journal = {Proceedings of the National Academy of Sciences},
	author = {Proekt, Alex and Banavar, Jayanth R. and Maritan, Amos and Pfaff, Donald W.},
	month = jun,
	year = {2012},
	pages = {10564--10569},
}

@article{dawid_present_1984,
	title = {Present {Position} and {Potential} {Developments}: {Some} {Personal} {Views}: {Statistical} {Theory}: {The} {Prequential} {Approach}},
	volume = {147},
	issn = {00359238},
	shorttitle = {Present {Position} and {Potential} {Developments}},
	url = {https://www.jstor.org/stable/10.2307/2981683?origin=crossref},
	doi = {10.2307/2981683},
	language = {en},
	number = {2},
	urldate = {2024-12-11},
	journal = {Journal of the Royal Statistical Society. Series A (General)},
	author = {Dawid, A. P.},
	year = {1984},
	pages = {278},
}

@article{vajna_modelling_2013,
	title = {Modelling bursty time series},
	volume = {15},
	issn = {1367-2630},
	url = {https://dx.doi.org/10.1088/1367-2630/15/10/103023},
	doi = {10.1088/1367-2630/15/10/103023},
	language = {en},
	number = {10},
	urldate = {2025-02-09},
	journal = {New Journal of Physics},
	author = {Vajna, Szabolcs and Tóth, Bálint and Kertész, János},
	month = oct,
	year = {2013},
	note = {Publisher: IOP Publishing},
	pages = {103023},
}

@book{cox_renewal_1967,
	series = {Methuen science paperbacks},
	title = {Renewal {Theory}},
	isbn = {978-0-412-20570-5},
	url = {https://books.google.com/books?id=KR4IAQAAIAAJ},
	publisher = {Taylor \& Francis},
	author = {Cox, D.R.},
	year = {1967},
	lccn = {lc68113847},
}

@article{common_scaling1,
  title = {Common scaling patterns in intertrade times of U. S. stocks},
  author = {Ivanov, Plamen Ch. and Yuen, Ainslie and Podobnik, Boris and Lee, Youngki},
  journal = {Phys. Rev. E},
  volume = {69},
  issue = {5},
  pages = {056107},
  numpages = {7},
  year = {2004},
  month = {May},
  publisher = {American Physical Society},
  doi = {10.1103/PhysRevE.69.056107},
  url = {https://link.aps.org/doi/10.1103/PhysRevE.69.056107}
}

@article{SONG2006527,
title = {Three types of power-law distribution of forest fires in Japan},
journal = {Ecological Modelling},
volume = {196},
number = {3},
pages = {527-532},
year = {2006},
issn = {0304-3800},
doi = {https://doi.org/10.1016/j.ecolmodel.2006.02.033},
url = {https://www.sciencedirect.com/science/article/pii/S0304380006000962},
author = {Weiguo Song and Jian Wang and Kohyu Satoh and Weicheng Fan}
}

@article{ogata1988,
 ISSN = {01621459, 1537274X},
 URL = {http://www.jstor.org/stable/2288914},
 author = {Yosihiko Ogata},
 journal = {Journal of the American Statistical Association},
 number = {401},
 pages = {9--27},
 publisher = {[American Statistical Association, Taylor & Francis, Ltd.]},
 title = {Statistical Models for Earthquake Occurrences and Residual Analysis for Point Processes},
 urldate = {2025-03-14},
 volume = {83},
 year = {1988}
}

@article{clauset2009,
author = {Clauset, Aaron and Shalizi, Cosma Rohilla and Newman, M. E. J.},
title = {Power-Law Distributions in Empirical Data},
journal = {SIAM Review},
volume = {51},
number = {4},
pages = {661-703},
year = {2009},
doi = {10.1137/070710111},

URL = { 
        https://doi.org/10.1137/070710111
},
eprint = { 
        https://doi.org/10.1137/070710111
}
}

@article{plerou2000,
  title = {Economic fluctuations and anomalous diffusion},
  author = {Plerou, Vasiliki and Gopikrishnan, Parameswaran and Nunes Amaral, Lu\'{\i}s A. and Gabaix, Xavier and Eugene Stanley, H.},
  journal = {Phys. Rev. E},
  volume = {62},
  issue = {3},
  pages = {R3023--R3026},
  numpages = {0},
  year = {2000},
  month = {Sep},
  publisher = {American Physical Society},
  doi = {10.1103/PhysRevE.62.R3023},
  url = {https://link.aps.org/doi/10.1103/PhysRevE.62.R3023}
}

@article{wheatland_understanding_2002,
	title = {Understanding {Solar} {Flare} {Waiting}-{Time} {Distributions}},
	volume = {211},
	issn = {1573-093X},
	url = {https://doi.org/10.1023/A:1022430308641},
	doi = {10.1023/A:1022430308641},
	number = {1},
	journal = {Solar Physics},
	author = {Wheatland, M.S. and Litvinenko, Y.E.},
	month = dec,
	year = {2002},
	pages = {255--274},
}

@article{HONG20076,
title = {Power law in firms bankruptcy},
journal = {Physics Letters A},
volume = {361},
number = {1},
pages = {6-8},
year = {2007},
issn = {0375-9601},
doi = {https://doi.org/10.1016/j.physleta.2006.09.034},
url = {https://www.sciencedirect.com/science/article/pii/S0375960106014046},
author = {Byoung Hee Hong and Kyoung Eun Lee and Jae Woo Lee}
}

@article{kesten1973,
author = {Harry Kesten},
title = {{Random difference equations and Renewal theory for products of random matrices}},
volume = {131},
journal = {Acta Mathematica},
number = {none},
publisher = {Institut Mittag-Leffler},
pages = {207 -- 248},
year = {1973},
doi = {10.1007/BF02392040},
URL = {https://doi.org/10.1007/BF02392040}
}

@article{Vazques2006,
  title = {Modeling bursts and heavy tails in human dynamics},
  author = {V\'azquez, Alexei and Oliveira, Jo\~ao Gama and Dezs\"o, Zolt\'an and Goh, Kwang-Il and Kondor, Imre and Barab\'asi, Albert-L\'aszl\'o},
  journal = {Phys. Rev. E},
  volume = {73},
  issue = {3},
  pages = {036127},
  numpages = {19},
  year = {2006},
  month = {Mar},
  publisher = {American Physical Society},
  doi = {10.1103/PhysRevE.73.036127},
  url = {https://link.aps.org/doi/10.1103/PhysRevE.73.036127}
}

@article{Jo2015,
  title = {Correlated bursts and the role of memory range},
  author = {Jo, Hang-Hyun and Perotti, Juan I. and Kaski, Kimmo and Kert\'esz, J\'anos},
  journal = {Phys. Rev. E},
  volume = {92},
  issue = {2},
  pages = {022814},
  numpages = {8},
  year = {2015},
  month = {Aug},
  publisher = {American Physical Society},
  doi = {10.1103/PhysRevE.92.022814},
  url = {https://link.aps.org/doi/10.1103/PhysRevE.92.022814}
}

@article{
Malmgren2008,
author = {R. Dean Malmgren  and Daniel B. Stouffer  and Adilson E. Motter  and Luís A. N. Amaral },
title = {A Poissonian explanation for heavy tails in e-mail communication},
journal = {Proceedings of the National Academy of Sciences},
volume = {105},
number = {47},
pages = {18153-18158},
year = {2008},
doi = {10.1073/pnas.0800332105},
URL = {https://www.pnas.org/doi/abs/10.1073/pnas.0800332105},
eprint = {https://www.pnas.org/doi/pdf/10.1073/pnas.0800332105}}

@article{Biro2005,
  title = {Power-Law Tails from Multiplicative Noise},
  author = {Bir\'o, Tam\'as S. and Jakov\'ac, Antal},
  journal = {Phys. Rev. Lett.},
  volume = {94},
  issue = {13},
  pages = {132302},
  numpages = {4},
  year = {2005},
  month = {Apr},
  publisher = {American Physical Society},
  doi = {10.1103/PhysRevLett.94.132302},
  url = {https://link.aps.org/doi/10.1103/PhysRevLett.94.132302}
}

@book{ross1995stochastic,
  title     = {Stochastic Processes},
  author    = {Sheldon M. Ross},
  edition   = {2},
  year      = {1995},
  publisher = {Wiley},
  address   = {New York},
  isbn      = {978-0-471-12062-9},
  pages     = {544}
}

@article{Bak1998,
  title = {Self-organized criticality},
  author = {Bak, Per and Tang, Chao and Wiesenfeld, Kurt},
  journal = {Phys. Rev. A},
  volume = {38},
  issue = {1},
  pages = {364--374},
  numpages = {0},
  year = {1988},
  month = {Jul},
  publisher = {American Physical Society},
  doi = {10.1103/PhysRevA.38.364},
  url = {https://link.aps.org/doi/10.1103/PhysRevA.38.364}
}

@article{Baiesi2006,
  title = {Intensity Thresholds and the Statistics of the Temporal Occurrence of Solar Flares},
  author = {Baiesi, Marco and Paczuski, Maya and Stella, Attilio L.},
  journal = {Phys. Rev. Lett.},
  volume = {96},
  issue = {5},
  pages = {051103},
  numpages = {4},
  year = {2006},
  month = {Feb},
  publisher = {American Physical Society},
  doi = {10.1103/PhysRevLett.96.051103},
  url = {https://link.aps.org/doi/10.1103/PhysRevLett.96.051103}
}

@article{Gandica_2017,
  title={Stationarity of the inter-event power-law distributions},
  author={Gandica, Yerali and Carvalho, João and dos Aidos, Fernando Sampaio and Lambiotte, Renaud and Carletti, Timoteo},
  journal={PLOS ONE},
  volume={12},
  number={3},
  pages={e0174509},
  year={2017},
  publisher={Public Library of Science},
  doi={10.1371/journal.pone.0174509},
  url={https://journals.plos.org/plosone/article?id=10.1371/journal.pone.0174509}
}

@article{Norman2001,
doi = {10.1086/321678},
url = {https://dx.doi.org/10.1086/321678},
year = {2001},
month = {aug},
publisher = {},
volume = {557},
number = {2},
pages = {891},
author = {Norman, James P. and Charbonneau, Paul and McIntosh, Scott W. and Liu, Han-Li},
title = {Waiting-Time Distributions in Lattice Models of Solar
Flares},
journal = {The Astrophysical Journal}
}

@article{Wheatland2000,
doi = {10.1086/312739},
url = {https://dx.doi.org/10.1086/312739},
year = {2000},
month = {jun},
publisher = {},
volume = {536},
number = {2},
pages = {L109},
author = {Wheatland, M. S.},
title = {The Origin of the Solar Flare Waiting-Time
Distribution},
journal = {The Astrophysical Journal}
}

@article{zhao_2011,
	title = {Empirical {Analysis} on the {Human} {Dynamics} of a {Large}-{Scale} {Short} {Message} {Communication} {System}},
	volume = {28},
	url = {https://dx.doi.org/10.1088/0256-307X/28/6/068901},
	doi = {10.1088/0256-307X/28/6/068901},
	number = {6},
	journal = {Chinese Physics Letters},
	author = {{and} and {and}},
	month = jun,
	year = {2011},
	pages = {068901},
}

@article{Sornette_2007,
author = {Saichev, Alexander and Sornette, Didier},
title = {Theory of earthquake recurrence times},
journal = {Journal of Geophysical Research: Solid Earth},
volume = {112},
number = {B4},
pages = {},
doi = {https://doi.org/10.1029/2006JB004536},
url = {https://agupubs.onlinelibrary.wiley.com/doi/abs/10.1029/2006JB004536},
eprint = {https://agupubs.onlinelibrary.wiley.com/doi/pdf/10.1029/2006JB004536},
year = {2007}
}

@article{Sornette_2006,
  title = {``Universal'' Distribution of Interearthquake Times Explained},
  author = {Saichev, A. and Sornette, D.},
  journal = {Phys. Rev. Lett.},
  volume = {97},
  issue = {7},
  pages = {078501},
  numpages = {4},
  year = {2006},
  month = {Aug},
  publisher = {American Physical Society},
  doi = {10.1103/PhysRevLett.97.078501},
  url = {https://link.aps.org/doi/10.1103/PhysRevLett.97.078501}
}

@article{Bacry_2015,
author = {Bacry, Emmanuel and Mastromatteo, Iacopo and Muzy, Jean-Fran\c{c}ois},
title = {Hawkes Processes in Finance},
journal = {Market Microstructure and Liquidity},
volume = {01},
number = {01},
pages = {1550005},
year = {2015},
doi = {10.1142/S2382626615500057},
URL = { 
        https://doi.org/10.1142/S2382626615500057
},
eprint = {    
        https://doi.org/10.1142/S2382626615500057
}
}

@book{Oksendal2003,
  author    = {Bernt {Ø}ksendal},
  title     = {Stochastic Differential Equations: An Introduction with Applications},
  series    = {Universitext},
  edition   = {6},
  year      = {2003},
  publisher = {Springer Berlin Heidelberg},
  address   = {Berlin, Heidelberg},
  isbn      = {978-3-540-04758-2},
  doi       = {10.1007/978-3-642-14394-6}
}

\end{document}